\documentclass[acmlarge]{acmart}

\AtBeginDocument{%
	\providecommand\BibTeX{{%
			\normalfont B\kern-0.5em{\scshape i\kern-0.25em b}\kern-0.8em\TeX}}}

\acmYear{2021}
\acmDOI{10.1145/3418501}

\setcopyright{acmcopyright}
\acmJournal{TOIT}
\acmYear{2021} \acmVolume{1} \acmNumber{1} \acmArticle{1} \acmMonth{4} \acmPrice{15.00}\acmDOI{10.1145/3418501}

\usepackage{adjustbox}
\usepackage{blkarray, bigstrut}
\usepackage{multirow} 

\usepackage[british,american]{babel}
\usepackage{array,multirow}
\usepackage{color}
\usepackage{cuted}
\usepackage{subcaption}
\usepackage{flushend}
\RequirePackage{graphicx}
\RequirePackage{mathptmx}      
\RequirePackage{flushend}
\usepackage{amsmath}
\usepackage{hyperref}
\usepackage{epstopdf}
\usepackage{wrapfig}
\usepackage{latexsym}
\usepackage{amsthm}
\usepackage{amsfonts}
\usepackage{amsmath} 
\usepackage{graphicx}
\usepackage{latexsym}
\usepackage{booktabs}

\usepackage{calc}
 \usepackage{enumitem}
\usepackage{mathptmx}

\usepackage{relsize}
\usepackage{stfloats}

\usepackage[style=base]{caption}
\usepackage{subcaption} 
\usepackage{breqn}
\usepackage{xcolor}

\usepackage{rotating}

\usepackage{amsthm}

\usepackage{colortbl}
\definecolor{LightCyan}{RGB}{255, 255, 100}
\definecolor{LightCyan1}{RGB}{162, 255, 100}

\definecolor{Gray}{gray}{0.9}
\newcommand{\coldscr}{\cellcolor{Gray}}

\usepackage{algorithmicx}
\usepackage{algorithm}

\usepackage{algpseudocode}
\usepackage{amsmath}
 \usepackage{tcolorbox}

\begin{document}

\title{Joint QoS-aware and Cost-efficient Task Scheduling for Fog-Cloud Resources in a Volunteer Computing System}

\author{Farooq Hoseiny}
\affiliation{%
	\institution{Department of Computer Engineering and IT, University of Kurdistan}
	\city{Sanandaj}
	\country{Iran}
}
\email{farooq.hoseiny@eng.uok.ac.ir}

\author{Sadoon Azizi}
\authornote{This is the corresponding author}
\authornote{All authors contributed equally to this research.}
\affiliation{%
	\institution{Department of Computer Engineering and IT, University of Kurdistan}
	\city{Sanandaj}
	\country{Iran}
}
\email{s.azizi@uok.ac.ir}

\author{Mohammad Shojafar}
\affiliation{%
	\institution{ICS/5GIC, University of Surrey}
	\streetaddress{1 Th{\o}rv{\"a}ld Circle}
	\city{Guildford}
	\country{United Kingdom}}
\email{m.shojafar@surrey.ac.uk}

\author{Rahim Tafazolli}
\affiliation{%
	\institution{ICS/5GIC, University of Surrey}
	\streetaddress{1 Th{\o}rv{\"a}ld Circle}
	\city{Guildford}
	\country{United Kingdom}}
\email{r.tafazolli@surrey.ac.uk}

\renewcommand{\shortauthors}{Azizi and Shojafar, et al.}

\begin{abstract}
 Volunteer computing is an Internet-based distributed computing system in which volunteers share their extra available resources to manage large-scale tasks. However, computing devices in a Volunteer Computing System (VCS) are highly dynamic and heterogeneous in terms of their processing power, monetary cost, and data transferring latency. To ensure both the high Quality of Service (QoS) and low cost for different requests, all of the available computing resources must be used efficiently. Task scheduling is an NP-hard problem that is considered one of the main critical challenges in a heterogeneous VCS. Due to this, in this paper, we design two task scheduling algorithms for VCSs, named \textit{Min-CCV} and \textit{Min-V}. The main goal of the proposed algorithms is jointly minimizing the computation, communication and delay violation cost for the Internet of Things (IoT) requests. Our extensive simulation results show that proposed algorithms are able to allocate tasks to volunteer fog/cloud resources more efficiently than the state-of-the-art. Specifically, our algorithms improve the deadline satisfaction task rates by around 99.5\% and decrease the total cost between 15 to 53\% in comparison with the genetic-based algorithm.
\end{abstract}



\maketitle

\section{Introduction}\label{sec:1.1}
 A Volunteer Computing System (VCS) is an opportunistic model that uses the extra available resources of volunteer devices to provide cheaper and greener infrastructures and services~\cite{anderson2005high,lee2010robust,ghafarian2015cloud}. These devices are characterized by their dynamic, distributed, and heterogeneous nature where through the use of them, many complex problems can be managed. Volunteer devices can be deployed in a fog or a cloud environment, where we call them Fog Nodes (FNs) or Cloud Node (CNs), respectively. Although FNs usually provide significant benefits in terms of response time and cost, they suffer from limited computing resources. As CNs offers more powerful resources, using the CNs along with the FNs provides a promising computing environment for efficient management of complex applications.

  The manifestation of the connection of devices like sensors, actuators, mobile/smartphones, and home/work appliances has led to many Internet of Things (IoT) applications. Many of these applications, such as connected and autonomous cars, augmented reality, industrial robotics, video surveillance, and real-time manufacturing, give rise to the need for low response time~\cite{1,2}. Other applications, such as big data analysis and machine learning, are latency tolerance. Most of IoT applications are composed of a series of independent tasks, also called Bag of Tasks (BoTs), with different input parameters~\cite{8}. The process of allocating available volunteer resources to a set of tasks is called task scheduling which is a critical challenge in the volunteer fog-cloud environment.

\subsection{Motivation}\label{sec:1.2}
The problem of scheduling tasks in volunteer fog-cloud computing systems significantly affects the Quality of Service (QoS) and the monetary cost~\cite{5}. In the real world, there are many latency-sensitive and latency-tolerant IoT applications with different requirements. This increases the complexity of scheduling and managing of them. Therefore, we require to present an efficient solution for task scheduling in this environment~\cite{6}.   


In the literature, several solutions has been proposed for solving task scheduling problem in fog-cloud domains~\cite{9,10,11,12,15,16,17,18,19,20,21,22,23,hassanpriority2020} which have some limitations as follows. A vast majority of them do not take into account the priority of tasks and assume the same priority for IoT tasks ~\cite{7,8,9,10,12}.  Most of the proposed approaches ignore the violation cost of the system in their problem formulation, except for a few of them~\cite{6,13}. In this work, we consider the priority of tasks based on their deadlines, violation cost and network latency for scheduling them in fog-cloud computing. To link the gaps to the contribution of the paper, some questions arise: i) Is it possible to present an efficient algorithm impose some scheduling policies to mitigate the cost and QoS of volunteer requests in fog-cloud environments? ii) Can we assure that the proposed methods could bring high QoS for real-time tasks in fog-cloud environments compared to the literature? And, iii) How can the collaboration of fog-cloud affect the QoS and monetary cost of volunteer requests?

\subsection{Goal and Contribution}\label{sec:1.3}

 In this study, we propose \textit{two} efficient heuristic scheduling algorithms for a batch of tasks, called \textit{Min-CCV}, (minimize \texttt{\textbf{C}}omputation, \texttt{\textbf{C}}ommunication and \texttt{\textbf{V}}iolation costs) and \textit{Min-V}, (minimize \texttt{\textbf{V}}iolation cost), that provide high QoS for IoT requests and low resources cost in a VCS. The main goal of Min-CCV is minimizing the computation, communication and violation cost of the system. However, Min-V aims to minimize the violation cost of the systems as far as possible. Both of the proposed algorithms aim to provide high QoS, in terms of the percentage of deadline satisfying tasks, and low cost in terms of computation and communication aspects.

In summary, the main contributions of the present research are as follows:

\begin{itemize}[leftmargin=*]
	\item We formulate the task scheduling problem considering volunteer requests in fog-cloud computing as a Mixed Integer Linear Programming (MILP) with the purposes of mutual supporting QoS for IoT tasks and low computation and communication cost for the fog-cloud system.
    \item To solve the above problem, we propose two efficient heuristic algorithms. The first one focuses on the minimization of computation, communication and violation costs, while the second one is appropriate for the minimization of the deadline violation cost.
    \item To evaluate the efficiency of the proposed algorithms in terms of QoS, makespan and total cost, we conduct extensive experiments and compare it against the state-of-the-art and confirm the effectiveness of our scheduling algorithms for IoT tasks in volunteer computing systems
\end{itemize}

\subsection{Organization}\label{sec:1.4}
The rest of the paper can be outlined as follows. In Section~\ref{pw}, we delineate the related works. In Section~\ref{approach-problemformulation}, we sketch the framework of the fog-cloud environment and formulate the task scheduling problem as a MILP. We present two heuristic algorithms which are presented in Section~\ref{proposedmethod}. Section~\ref{simulation} describes the presented scenarios, testing methods used to comparisons and describe the simulation results. Section~\ref{discuss} gives some discussions and limitations regarding the proposed policies. Finally, In Section~\ref{conc}, we summarize the achievements and concludes the paper.

\section{Related Work}\label{pw}
 
 In this section, we explain the existing task scheduling algorithms applied in the volunteer fog and cloud systems. Most of the static task scheduling approaches are divided into two categories: \textit{heuristic-based} and \textit{metaheuristic-based} algorithms. In the rest of the section, we briefly review some various algorithms. The algorithms studied here have all been tested in heterogeneous volunteer systems. The heterogeneous volunteer system consists of nodes with different computations and communication attributes that can be consist of both layer, fog, and cloud layers. 
 
 \subsection{Heuristic-based Fog algorithms}\label{heuristic-fog-algorithms}
 
 Heuristic methods are trying to return a suitable solution in the shortest possible time to reduce delay in the fog-cloud environment. Pham et al. \cite{16} propose a cost- and makespan-aware workflow scheduling algorithm in the fog-cloud computing system. The algorithm has three phases: task prioritizing phase which specifies the priority level of each task, node selection phase in which the most suitable node for executing each task is selected, and task reassignment phase to improve the QoS of the system. Liu et al. \cite{liu2018task} introduce an improved classification mining algorithm based on association rules and combine it with the task scheduling process in fog computing. The main objective of this study is reducing the execution time and the average waiting time of the scheduled tasks. Choudhari et al. \cite{choudhari2018prioritized} propose a priority-aware task scheduling algorithm to provide low response time for client requests. Upon receiving an incoming client request, the nearest fog server manager determines its priority level and checks whether its fog nodes can handle it or not. If so, the request is processed, and the output is passed to the client. Otherwise, if its required resources can be satisfied by the other fog servers, it is sent to the one or more of them. In the case of insufficient resources in the fog domain, the request is moved to the cloud. To minimize service delay in a heterogeneous fog environment, Liu et al. \cite{liu2018dats} propose a dispersive stable task scheduling algorithm. The algorithm consists of two major phases. In the first phase (i.e., the computing resources competition), initially, the processing efficiency of computing nodes is calculated, and then a pairwise stable matching between task nodes and computing nodes is obtained. In the second phase (i.e., tasks assignment), the synchronized task scheduling algorithm is called to the optimal decision-making process.
 
 Auluck et al. \cite{20} introduce two heuristic algorithms for real-time task scheduling generated by autonomous cars. In their study, the authors consider an embedded-fog-cloud framework and three types of tasks consist of hard, firm and soft ones. Their proposed algorithms assign the application tasks to the most appropriate processors, ensuring that the overall communication delay is minimized. In \cite{10}, Zhang et al. investigate the task scheduling problem in the voluntary-mode fog networks. They construct a general analytical model for the problem and propose a delay-optimal task scheduling algorithm to reduce the overall task processing delay. Stavrinides and Karatza \cite{17} present a heuristic approach to scheduling multiple real-time workflows in fog and cloud systems. In this work, the fog nodes are responsible for executing the communication-intensive tasks with low computation requirements while computation-intensive tasks with low communication demands are submitted to the cloud nodes. Moreover, during the scheduling process, tasks are prioritized according to the deadline of their job. Then, each task is assigned to the VM that provides the earliest estimated finish time. Benblidia et al. in \cite{9} propose a fuzzy logic approach for scheduling tasks in fog-cloud computing. The authors rank fog nodes according to both user preferences and the features of fog nodes.
  \begin{table*}[!htp]
\centering
\caption{Comparison of related works. BoT:= Bag of Tasks; CPS:= Cyber Physical System.}\label{relatedcomp}
\begin{adjustbox}{max width=\textwidth}
\begin{tabular}{lllccccl}
\hline
\textbf{Reference}&\textbf{System model}&\textbf{Application}&\textbf{Resource}&\textbf{Response}&\textbf{Deadline-} &\textbf{Runtime}&\textbf{Suggested for}\\
&&\textbf{type}&\textbf{cost}&\textbf{time}& \textbf{aware}&&\\
\hline
\textbf{Heuristic}&&&&&&&\\
Pham et al. \cite{16}&Fog-cloud&Workflow&
$\checkmark$&$\checkmark$&$\checkmark$&Low&Generic\\
Liu et al. \cite{liu2018task}&Fog&BoT&$\times$&$\checkmark$&
$\times$&Moderate&Generic\\
Choudhari et al. \cite{choudhari2018prioritized}&Fog-cloud&BoT&$\times$&\checkmark&$\checkmark$&
Low&Generic\\
Liu et al. \cite{liu2018dats}&Fog-cloud&BoT&$\times$&$\checkmark$&$\times$&Low&Heterogeneous fog networks\\
Auluck et al. \cite{20}&Fog-cloud&BoT&
$\times$&$\checkmark$&$\checkmark$&Low&Autonomous cars\\
Zhang et al. \cite{10}&Fog&BoT&
$\times$&$\checkmark$&$\times$&Low&Voluntary-mode fog networks\\
Stavrinides et al. \cite{17}&Fog-cloud&Workflow&
$\times$&$\checkmark$&$\checkmark$&Low&Real-time IoT workflows\\
Benblidia et al. \cite{9}&Fog-cloud&BoT&
$\checkmark$&$\checkmark$&$\times$&Moderate&Generic\\
\hline
\textbf{Min-CCV (This study)} &\textbf{Fog-cloud}&\textbf{BoT}&\textbf{$\checkmark$}&\textbf{$\checkmark$}&\textbf{$\checkmark$}&\textbf{Very Low}&\textbf{Generic}\\
\textbf{Min-V (This study)} &\textbf{Fog-cloud}&\textbf{BoT}&
\textbf{$\checkmark$}&\textbf{$\checkmark$}&\textbf{$\checkmark$}&\textbf{Low}&\textbf{Generic}\\
\hline
\textbf{Metaheuristic}&&&&&&&\\
Bitam et al. \cite{21}&Fog&BoT&$\checkmark$&$\checkmark$&
$\times$&High&Mobile users\\
Mishra et al. \cite{19}&Fog&BoT&$\times$&$\checkmark$&
$\times$&High&Industrial applications\\
Gill et al. \cite{b4}&Fog-cloud&Workflow&$\times$&$\checkmark$&
$\times$&High&Smart homes\\
Nguyen et al. \cite{8}&Fog-cloud&BoT&$\checkmark$&$\checkmark$&
$\times$&High&Generic\\
Aburukba et al. \cite{15}&Fog-cloud&Workflow&$\times$&$\checkmark$&
$\checkmark$&Low&Generic\\
Wang et al. \cite{11}&Fog-cloud&BoT&$\times$&$\checkmark$&
$\times$&High&Smart factories\\
Ghobaei-Arani et al. \cite{12}&Fog&BoT&$\times$&$\checkmark$&
$\times$&High&CPS applications\\
Javanmardi et al. \cite{22}&Fog-cloud&BoT&$\times$&$\checkmark$&
$\times$&High&Mobile IoT\\
\hline
\end{tabular}
\end{adjustbox}
\end{table*}
   \subsection{Metaheuristic-based Fog algorithms}\label{metaheuristic-fog-algorithms}
   Metaheuristic algorithms perform a random search to find a reasonable solution to an optimization scheduling problem~\cite{14}. Bitam et al. \cite{21} deal with the job scheduling problem in a fog environment. This study aims to find a good trade-off between two performance metrics, i.e., the time of execution and the memory allocated. To achieve this goal, the authors propose a bees life algorithm as a swarm-based optimization approach. In \cite{19}, Mishra et al. formulate the scheduling of service requests as a bi-objective optimization problem to minimize the makespan and energy consumption. Then, they solve the problem using three metaheuristic algorithms; particle swarm optimization (PSO), binary PSO (BPSO) and bat algorithm (BAT). Gill et al. \cite{b4} propose a new resource management technique which emphasizes on task scheduling for fog-enabled cloud computing. They leverage PSO to reduce response time, latency, network bandwidth, and the consumption of energy simultaneously. This work implements these four QoS parameters in the PSO fitness function and uses predefined weights to prioritize them. Nguyen et al. \cite{8} propose a genetic-based algorithm for the scheduling of tasks in fog-cloud computing to reduce the makespan and cost of computation, storage, and communication of resources. Similarly, Aburukba et al. in \cite{15} model the task scheduling as an integer linear programming problem and propose a customized genetic algorithm to minimize the overall service request latency.
   
   To reduce the delay and energy consumption, Wang and Li \cite{11} combine the advantages of the particle swarm optimization and ant colony optimization and propose a hybrid heuristic approach to scheduling tasks in smart factory environments. Ghobaei-Arani et al. in \cite{12} investigate the scheduling of tasks in the fog environment with a focus on cyber-physical system (CPS) applications. They take into account the execution and transfer time of tasks as objective functions and present a moth-flame optimization algorithm to provide high QoS for CPS applications. Javanmardi et al. \cite{22} propose a new joint meta-heuristic method called FPFTS combining PSO and Fuzzy methods to tackle the fog task scheduling problem. To do this, the authors implement a three-layered architecture which includes IoT devices, fog nodes instantiated on the edge layer and upper layer, which is a cloud data center and their transparent cloud service providers. Moreover, they give their tested scenarios enforced on mobility devices and fog device characteristics and their communications methods.
   
   Table \ref{relatedcomp} presents an overview of the related works and highlights the difference between this study and present works. Although previous works have considered some aspects of the task scheduling problem, there is still room for further improvement in terms of the problem formulation and the efficiency of algorithms. Hence, in this study, we take into account the computation and communication cost profile of fog and cloud nodes and the response time of tasks concerning their deadlines for Bag-of-Tasks (BoT) optimization problem in fog-cloud environments and finally we generate two efficient heuristic algorithms to solve BoT.

\section{Proposed Architecture and Problem Formulation}\label{approach-problemformulation}
In this section, we first explain the proposed architecture and the components participate in the tasks scheduling process in a VCS. Then we formulate the task scheduling problem as a MILP. 

\subsection{Proposed Architecture}\label{architecture}
\begin{figure}[!hbpt]
	\centering
    	\centering
    	\includegraphics[width=0.7\linewidth]{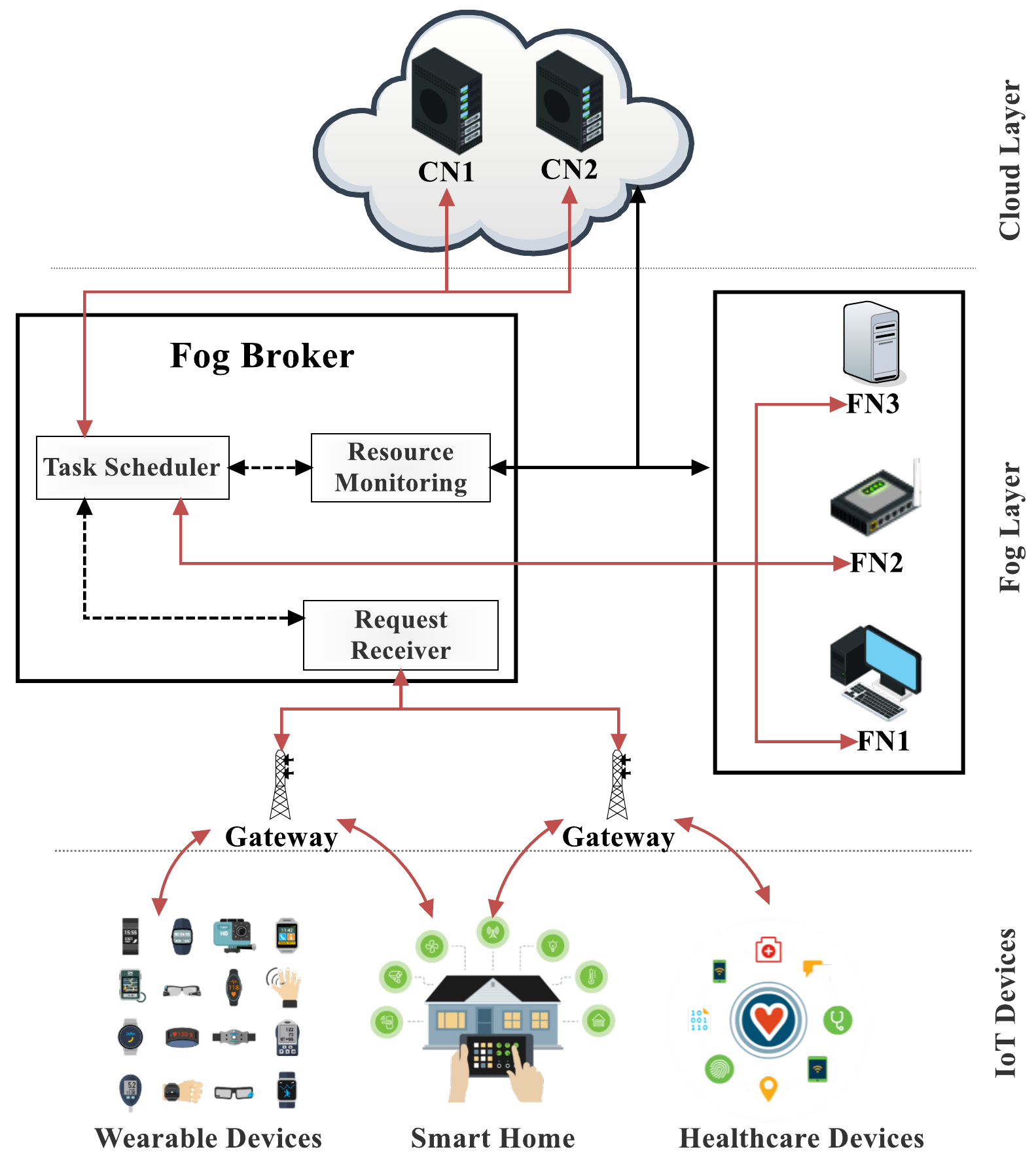}
    	\caption{The considered architecture for the proposed volunteer computing system. FN:= Fog Node, CN:= Cloud Node.}
    	\label{fig:Fog-cloud-architecture}
   \end{figure}

 Fig.~\ref{fig:Fog-cloud-architecture} presents our proposed architecture for a VCS. It consists of \textit{three} layers, namely, IoT devices, fog, and cloud nodes. The first layer includes heterogeneous IoT devices like wearable devices, smart home sensors, and healthcare devices. These devices can request some demands formed as jobs that are sent to the higher layers to run on fog/cloud nodes. The second layer consists of computers, mini-servers, routers, access points and other nodes called FNs; each node has the capability of computing, networking, and storage. In the cloud layer, there exist some powerful volunteer servers with high computation and communication cost with a long distance from user IoT devices. If the request requires massive processing power, it should be sent to the CNs as they offer higher computing capacities than FNs.

 The main part of the proposed architecture is an entity called \textit{Fog Broker} which is deployed in the fog layer. Fog Broker consists of three major components that are \textit{Request Receiver}, \textit{Resource Monitoring} and \textit{Task Scheduler}. \textit{Request Receiver} receives all requests from the distributed IoT devices through gateways. This component first validates the integrity and correctness of the request. Then, it estimates all of the relevant parameters of a submitted request such as the number of tasks and characteristics of each task. After that, it sends requests to the task scheduler. \textit{Resource Monitoring} is a unit that is responsible for periodically collecting and monitoring the available resources of volunteer FNs and CNs. It shares the status of the nodes to the task scheduler to help the scheduling of the requests efficiently. \textit{Task Scheduler} is the heart of Fog Broker in which the proposed task scheduling algorithms are run. Based on the profiles of submitted tasks and available resource capacity of FNs and CNs, Task Scheduler solves an instance of the task scheduling problem in each time period and maps tasks to the nodes.

	 \begin{figure}
	 	\centering
	 	\includegraphics[width=0.75\linewidth]{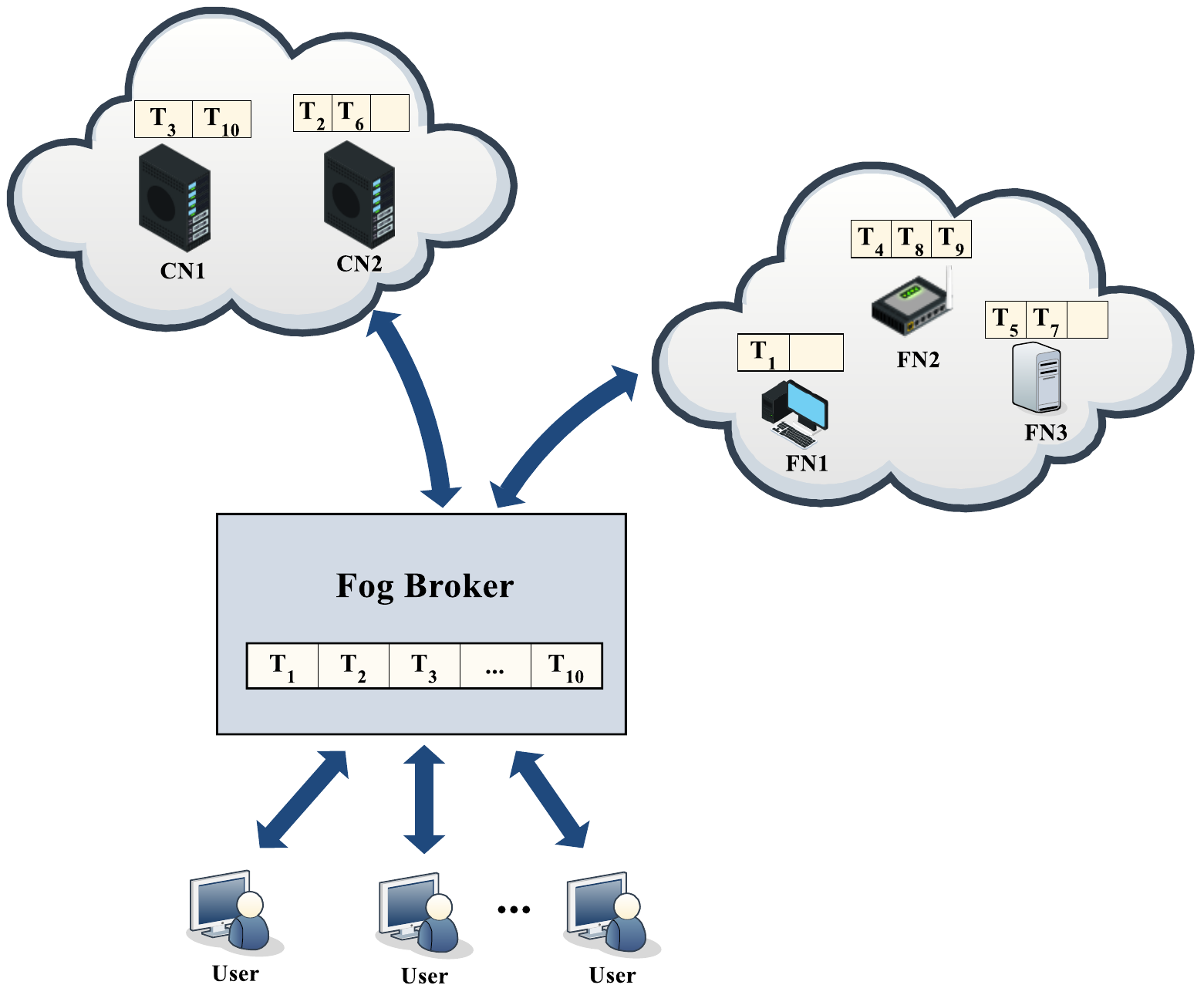}
	 	\caption{The proposed task scheduling approach. CN:= Cloud Node; FN:= Fog Node; $T_x$ = $x$-th task.}
	 	\label{fig:task-scheduling-architecture}
\end{figure}

 Fig.~\ref{fig:task-scheduling-architecture} demonstrates an overview of the task scheduling process in the proposed architecture. As we can see from the figure, IoT users send their requests to Fog Broker (see the shaded rectangle in the middle of Fig.~\ref{fig:task-scheduling-architecture}). Fog Broker decomposes a vector of independent tasks and creates a BoT list (see the vector of tasks inside the Fog Broker component in Fig.~\ref{fig:task-scheduling-architecture}). Depending on the embedded scheduling algorithm, it may sort them or not. Fog Broker is responsible for dispatching the BoT list to the available CNs and FNs in the network. Thus, Fog Broker allocates the tasks to the suitable FNs and CNs. If the tasks assigned to the CNs (see top-left cloud shape component in Fig.~\ref{fig:task-scheduling-architecture}) the cloud system will dispatch the tasks to the available server which is associated with the CN. Depends on the requirements of the tasks like delay-intensive and computation-intensive applications, the tasks can be assigned to the FNs (see top-right cloud shape component in Fig.~\ref{fig:task-scheduling-architecture}). FNs are nearer to the users can be the first choice of Fog Broker assignments. It helps to increase the throughput of the network by allocating the delay-sensitive applications on the FNs. All in all, the results of the processing of the task, whether located in CNs or FNs are obtained and sent to the Fog broker, and it can dispatch them to the corresponding users.

\begin{table}
	\begin{center}
		\caption{Symbol notations. ms:= millisecond; MI:= Million Instruction; MB:= Mega Byte; MIPS:= Million Instruction Per Second; G\$:= Grid Dollar; s:= second. }\label{notation}
		\footnotesize
		\begin{adjustbox}{max width=\textwidth}
			\label{tab:tab1}
			\begin{tabular}{|p{0.2cm}|c|p{7.5cm}|p{2.5cm}|p{3.2cm}|}
				\hline
				\coldscr & \coldscr \textbf{Symbol} & \coldscr \textbf{Definition} & \coldscr \textbf{Type - Unit} & \coldscr \textbf{calculates in Eq.} \\
				\hline
				\multirow{4.5}{0.5cm}{\begin{sideways}\textbf{Set}\end{sideways}}& $\pmb{T}$ & Set of independent tasks, where $\mid \pmb{T}\mid= n$&-&-\\
				&$\pmb{N}$ & Set of nodes, where $\mid \pmb{N}\mid= m$&-&-\\ 
				&$N_f$ & Set of fog nodes, where $\mid N_f\mid= f$&-&-\\ 
				&$N_c$ & Set of cloud nodes, where $\mid N_c\mid= c$&-&-\\ 
				\hline \hline
				\multirow{2.5}{0.5cm}{\begin{sideways}\textbf{Ind.}\end{sideways}}& $i$ & Index of tasks, $i\in \pmb{T}$&Integer - [units]&-\\
				&$j$ & Index of nodes, $j\in \pmb{N}$&Integer - [units]&-\\ 
				\hline \hline
				&$E_{ij}$& Execution time of $T_i$ on $N_j$&Continuous - [ms]&\eqref{eq:eq2}, \eqref{eq:eq3}, \eqref{eq:eq6-1}\\
				\multirow{11}{0.5cm}{\begin{sideways}\textbf{Input Parameters}\end{sideways}} 
				&$W_{ij}$& Waiting time of $T_i$ in the queue of $N_j$&Continuous - [ms]&\eqref{eq:eq6-1}\\
				&$T_i^s$& Number of instructions of $T_i$& Integer - [MI]&\eqref{eq:eq2}\\
				&$T_i^{mem}$& Required memory of $T_i$& Continuous - [MB]&\eqref{eq:eq3},\eqref{eq:eq11}\\
				&$T_i^{bw}$& Required bandwidth of $T_i$& Continuous - [MB]&\eqref{eq:eq5}\\
				&$T_i^{deadline}$& Required deadline time of $T_i$& Continuous - [ms]&\eqref{eq:eq7}\\
				&$T_i^{qos}$& Required quality of service of $T_i$& Continuous - [\%]&\eqref{eq:eq8}\\
				&$T_i^{response}$& Response time of $T_i$& Continuous - [ms]&\eqref{eq:eq6-1}, \eqref{eq:eq7}\\
				&$T_i^{penalty}$& Penalty rate of $T_i$& Continuous - [per \%]&\eqref{eq:eq8}\\
				&$N_j^{cpu}$& CPU processing rate of $N_j$& Integer - [MIPS]&\eqref{eq:eq2}\\
				&$N_j^{mem}$& Memory capacity of $N_j$& Continuous - [MB]&\eqref{eq:eq11}\\
				&$N_j^{delay}$& Delay between the task scheduler module and $N_j$& Continuous - [ms]&\eqref{eq:eq6-1}\\
				\hline\hline
				
			\multirow{3}{0.5cm}{\begin{sideways}\textbf{Constant}\end{sideways}}
			& $c_j^{p}$ & CPU usage cost of $N_j$ &Continuous - [G\$/s]&\eqref{eq:eq3}\\
			& $c_j^{m}$ & Memory usage cost of $N_j$ &Continuous - [G\$/MB]&\eqref{eq:eq3}\\
			& $c_j^{b}$ & Bandwidth usage cost of $N_j$ &Continuous - [G\$/MB]&\eqref{eq:eq5}\\
				\hline \hline
				\multirow{9}{0.5cm}{\begin{sideways}\textbf{Variables}\end{sideways}} 
				& $\mathbf{X}_{n\times m}$ & Allocation matrix &Binary - [units]&- \\ 
				& $x_{ij}$ & Decision variable showing if $T_i$ is allocated to $N_j$&Binary - [units]&\eqref{eq:eq1}, \eqref{eq:eq3}, \eqref{eq:eq5}, \eqref{eq:eq6-1}, \eqref{eq:eq11}, \eqref{eq:eq12} \\ 
				&${C}_i^{comp}$ & Computing cost for $T_i$ &Continuous - [G\$]&\eqref{eq:eq3}, \eqref{eq:eq4} \\
				&${C}_i^{comm}$ & Communication cost for $T_i$ &Continuous - [G\$]&\eqref{eq:eq5}, \eqref{eq:eq6} \\
				&${C}_i^{viol}$ & Violation cost for $T_i$ &Continuous - [G\$]&\eqref{eq:eq8}, \eqref{eq:eq8-1} \\
				&${V}_{i}$ & Violation time of $T_i$&Continuous - [\%]&\eqref{eq:eq7}, \eqref{eq:eq8} \\
				&$\mathcal{C}^{comp}$ & Total computing cost&Continuous - [G\$]&\eqref{eq:eq4}, \eqref{eq:eq9} \\
				&$\mathcal{C}^{comm}$ & Total communication cost&Continuous - [G\$]&\eqref{eq:eq6}, \eqref{eq:eq9} \\
				&$\mathcal{C}^{viol}$ & Total violation cost&Continuous - [G\$]&\eqref{eq:eq8-1}, \eqref{eq:eq9} \\
				&$\mathbb{C}^{tot}$ & Total cost  &Continuous - [G\$]&\eqref{eq:eq9}, \eqref{eq:eq10}\\
				\hline
			\end{tabular}
		\end{adjustbox}
	\end{center}
\end{table}
\subsection{Problem Formulation}\label{problemformulation}

In this part, we present the mathematical formulation for the task scheduling problem. Table~\ref{notation} shows the main notation used in this paper. 

 Assuming that a set of $n$ independent tasks submitted to the Task Scheduler module, as follows: $\pmb{T}=\{T_1,\ T_2,\ T_3, \ldots,\ T_n\}$. Each task $T_i$ has its attributes, such as a number of instructions, memory requirement, input/output file size, deadline, and QoS. Also, let a set of $m$ computing nodes including  $f$ FNs and $c$ CNs in the considered VCS where $\pmb{N} \triangleq N_f \cup N_c$ is expressed as: $\pmb{N}=\{N_1,\ N_2,\ N_3, \ldots,\ N_m\}$. Each node $N_j$ has properties such as the rate of CPU processing, cost of CPU usage, cost of memory usage, cost of bandwidth usage, size of memory, and delay of communication. Although CNs typically offer higher computation power than FNs, the monetary cost and communication latency are higher in the former.

Let $\mathbf{X}_{n\times m}$ be an allocation matrix to represent the status of tasks. If task $T_i$ is allocated to $N_j$, $x_{ij}$ equals 1; otherwise equals 0. We have

\begin{equation}
			\label{eq:eq1}
			x_{ij}=
			\begin{cases}
				1& \text{if $T_i$ is allocated to $N_j$}\\
				0& \text{otherwise}\\
			\end{cases},\quad
			\forall T_i\in \pmb{T}, \forall N_j\in \pmb{N}
\end{equation}

When a task is allocated to a node, it takes some time for that task to execute on the respective node. Therefore, we define $E_{ij}$ as the execution time of $T_i$ when it is allocated on $N_j$, which can be obtained using the following equation. 

\begin{equation}
\label{eq:eq2}
\centering 
E_{ij}=\frac{T_i^{s}}{N_j^{cpu}},\quad
			\forall T_i\in \pmb{T}, \forall N_j\in \pmb{N}
\end{equation}
in which $T_i^{s}$ and $N_j^{cpu}$ are the number of instruction of $T_i$ and the CPU processing rate of $N_j$, respectively.

 We next model the cost of a VCS, including the computation, communication and deadline violation cost.

\subsubsection{Computation Cost:}
As one computing node processes a task, a monetary cost must be paid. The computation cost of a given task consists of two parts: processing and memory cost that can be calculated as follows.

	\begin{equation}
		\label{eq:eq3}
		\centering 
		C_i^{comp}= \sum_{j=1}^{m}\left(c_j^p\times E_{ij}+c_j^m\times T_i^{mem}\right)\times x_{ij},\quad
		\forall i \in \{1,\ \ldots,\ n\}
	\end{equation}
where $T_i^{mem}$ depicts the required memory of task $T_i$. Also, $c_j^p$ and $c_j^m$ are constant numbers which respectively represent the CPU usage cost and memory usage cost of node $N_j$. Based on this equation, the total computing cost for a set of $n$ tasks is defined as below. 

\begin{equation}
			\label{eq:eq4}
			\centering 
			\mathcal{C}^{comp}= \sum_{i=1}^{n}C_i^{comp}.
\end{equation}

\subsubsection{Communication Cost:}

For a given task, in addition to the computation cost, the communication cost is also introduced \cite{8,16}. This cost depends on the sum of input and output data file sizes of that task and the cost of bandwidth usage per data unit of nodes. So let $T_i^{bw}$ be the amount of bandwidth requirement of task $T_i$ and $c_j^b$ be the cost of bandwidth usage per data unit of node $N_j$. The communication cost for task $T_i$ is obtained as follow.

\begin{equation}
		\label{eq:eq5}
		\centering 
		C_i^{comm}= \sum_{j=1}^{m} \left(c_j^b\times T_i^{bw}\right)\times x_{ij},\quad
		\forall i \in \{1,\ \ldots,\ n\} 
\end{equation}

Hence, the total communication cost for all of $n$ tasks is given by the following equation.

\begin{equation}
			\label{eq:eq6}
			\centering 
			\mathcal{C}^{comm}= \sum_{i=1}^{n}C_i^{comm}.
\end{equation}

\subsubsection{Deadline Violation Cost:}

To model the QoS, we need to present an appropriate formula for deadline violation cost. To this end, we first define the system response time for a given task $T_i$.  Here we define it as the time interval between the moment that the task scheduler module receives the input file and the moment it gets the output file. Hence, we should consider the delay between the task scheduler module and node $N_j$, depicted by $N_j^{delay}$, the execution time of task $T_i$ on node $N_j$,  see eq.~\eqref{eq:eq2}, and the waiting time of $T_i$ in the queue of $N_j$, denoted by $W_{ij}$. Take into account these parameters, the response time for task $T_i$ is obtained using the following equation.  

\begin{equation}
\label{eq:eq6-1}
\centering 
        T_i^{response}=\sum_{j=1}^{m}(2 \times N_j^{delay} + E_{ij}+W_{ij})\times x_{ij},  \quad
		                       	\forall i \in \{1,\ \ldots,\ n\}
\end{equation}

To measure the deadline violation cost for a given task $T_i$, we first define $V_{i}$ as the percentage of the violation. Here, we propose the following formula to obtain its value.

\begin{equation}
\label{eq:eq7}
\centering 
V_{i}=\frac{\max(0,T_i^{response}-T_i^{deadline})}{T_i^{deadline}} \times 100, \quad
		    	\forall i \in \{1,\ \ldots,\ n\}
\end{equation}
 where the range of $V_i$ is $[0,\infty]$. It is zero if the task response time is lower than deadline time (i.e., $T_i^{deadline}\geq T_i^{response}$). However, it can be can any positive value depending on the distance between response and deadline time.

To determine the violation cost for a given task $T_i$, we need its QoS requirement, $T_i^{qos}$, and the penalty that the Fog Broker must pay by one percent of the delay violation, $T_i^{penalty}$. It is worth to mention that a similar metric is defined in \cite{yousefpour2019fogplan}. 

\begin{equation}
\label{eq:eq8}
\centering 
C_i^{viol}= \left(V_i-\left(100-T_i^{qos}\right)\right)\times T_i^{penalty}, \quad
			\forall i \in \{1,\ \ldots,\ n\}, 
\end{equation}

Considering the above equation, we can calculate the deadline violation cost for the system as follows.

\begin{equation}
			\label{eq:eq8-1}
			\centering 
			\mathcal{C}^{viol}= \sum_{i=1}^{n}C_i^{viol}.
\end{equation}

Now we can obtain the total cost of a VCS using the following equation.

\begin{equation}
\label{eq:eq9}
\centering 
\mathbb{C}^{tot}= \mathcal{C}^{comp}\ + \ \mathcal{C}^{comm}\ + \  \mathcal{C}^{viol}
\end{equation}

\subsubsection{Overall Problem Formulation:}\label{overall-problem}
		
Our main objective is solving the task scheduling problem in a heterogeneous volunteer computing system in such a way that the total cost is minimized. Therefore, the final optimization formula is defined as the following MILP model.
		\begin{align}
			\label{eq:eq10}
			\centering 
			\min\ & \mathbb{C}^{tot} 
		\end{align}
		subject to:
		\begin{align}
			\centering 
			\label{eq:eq11}&\text{Memory constraint:}\quad x_{ij}\times T_i^{mem}\leq N_j^{mem},\quad \forall i \in \{1,\ \ldots,\ n\}, \forall j \in \{1,\ \ldots,\ m\} \\
			\label{eq:eq12}&\text{Task constraint:}\quad \sum_{j=1}^{m}x_{ij} = 1,\quad \forall i \in \{1,\ \ldots,\ n\} 
		\end{align}
		under binary control (decision) variable: $x_{ij}\in \{0,1\}$, $\forall i \in \{1,\ \ldots,\ n\}, \forall j \in \{1,\ \ldots,\ m\}$. 
		
To solve this MILP problem, we require to ensure that we have a task scheduling problem such that $n$ independent tasks want to be processed through $m$ nodes. Since the requirement of tasks and properties of nodes are different, there exist $m^n$ solutions for mapping the tasks to the nodes. In a large scale problem, $n$ and $m$ usually are on the scale of thousands~\cite{chen2013intelligent,bari2012data}, proposing an efficient algorithm is a challenging task. In the following section, we propose \textit{two} heuristic algorithms to find a suitable solution in a real-time, even for very large scale $n$s and $m$s.

\section{Proposed Algorithms}\label{proposedmethod}

In this part, we first introduce our heuristic algorithms for scheduling tasks in a VCS (see subsections~\ref{min-ccv} and ~\ref{min-v}). Then, we give an illustrative example to show how our proposed algorithms work (see subsection~\ref{proposedmethod:example}). Finally, we provide the time and space complexity analysis of the algorithms (see subsection~\ref{proposedmethod:complexity}).

\subsection{Min-CCV Algorithm}\label{min-ccv}
Min-CCV is a computation-, communication- and violation-aware task scheduling algorithm. Min-CCV algorithm allocates a task to a node which provides the lowest total cost, see eq.~\eqref{eq:eq10}, for that task. It is worth mentioning that the algorithm searches among all fog and cloud nodes that provide enough memory for the task,  see eq.~\eqref{eq:eq11}.

\begin{center}
\alglanguage{pseudocode}
 \begin{algorithm}
  \allowdisplaybreaks
  \caption{Min-CCV: computation, communication and violation-aware algorithm}
  \label{alg:Min-CCV}
  \begin{algorithmic}[1]
   \Procedure{Min-CCV}{$\pmb{T}$, $\pmb{N}$}
   \State{$availableTime [|\pmb{N}|]\leftarrow 0$}
   \For {\textbf{all} $T_i \in \pmb{T}$}\label{alg:start-tasks}
   \State{$C^{min} \leftarrow \infty$}
\For {\textbf{all} $N_j \in \pmb{N}$}\label{alg:start-nodes}
   \If{$T_i^{mem}\leq N_j^{mem}$}\label{alg:start-if-big}
    \State {calculate $C_i^{comp}$ using eq. \eqref{eq:eq3}}\label{alg:start-calculate}
    \State {calculate $C_i^{comm}$ using eq. \eqref{eq:eq5}}
    \State {calculate $C_i^{viol}$ using eq. \eqref{eq:eq7}}
    \State{$C^{tot} \leftarrow C_j^{comp}+C_j^{comm}+C_j^{viol}$}\label{alg:end-calculate}
    \If{$C^{tot} < C^{min}$}\label{alg:start-if-ctot}
    \State {$C^{min} \leftarrow C^{tot}$}
    \State {$index \leftarrow j$}
   \EndIf \label{alg:end-if-ctot}
   \EndIf
    \EndFor\label{alg:stop-nodes}
    \State {allocate $T_i$ to $N_{index}$}\label{alg:start-allocate}
    \State{$availableTime [N_{index}] \leftarrow availableTime [N_{index}] + E_{i,index}$}\label{alg:end-allocate}
   \EndFor\label{alg:stop-tasks}
   \EndProcedure
  \end{algorithmic}
 \end{algorithm}
\end{center}

Algorithm~\ref{alg:Min-CCV} presents the pseudo-code of the Min-CCV algorithm. It initially sets the available time of all nodes to 0 (line 2). Then, in the main loop (lines~\ref{alg:start-tasks} to~\ref{alg:stop-tasks}) it searches among all nodes and tries to find a node which can provides the least total cost including computation, communication and violation costs for a selected task (lines \ref{alg:start-nodes} to~\ref{alg:stop-nodes}). To this end, it first investigates the memory constraint (line~\ref{alg:start-if-big}) to check whether a node has enough memory or not. If so, it calculates the total cost for that node (lines~\ref{alg:start-calculate} to~\ref{alg:end-calculate}). After that, the node with the least cost is selected for a given task (lines~\ref{alg:start-if-ctot} to~\ref{alg:end-if-ctot}). In the end, we assign the task to the indexed node and update the available time of the indexed node (lines~\ref{alg:start-allocate} and~\ref{alg:end-allocate}).  

\subsection{Min-V Algorithm}\label{min-v}
Here we propose an efficient heuristic algorithm, called Min-V, for the batch mode in which a batch of tasks arrives at the task scheduler module. The corresponding task scheduling algorithm is executed. The main goal of Min-V is minimizing the delay violations as far as possible, i.e., the algorithm gives higher priority to QoS compared to cost. To achieve this, we first sort tasks in ascending order by their predefined deadline. Then, for each task, we create a set of FNs and CNs which meet the memory constraint and satisfy the task deadline. After that, we select the node with the minimum computation and communication cost. If there exist no nodes to fulfil the deadline requirement of a given task, we choose the one which offers the minimum violation cost.

\begin{center}
\alglanguage{pseudocode}
 \begin{algorithm}
  \allowdisplaybreaks
  \caption{Min-V: violation-aware algorithm}
  \label{alg:Min-V}
  \begin{algorithmic}[1]
   \Procedure{Min-V}{$\pmb{T}$, $\pmb{N}$}
    \State{$activeTime [|\pmb{N}|]\leftarrow 0$}
    \State{$AscSort (\pmb{T},T_i^{deadline})\quad \forall i\in \{1,\ldots,n\}$}\label{alg2:ascsort}
    \For {\textbf{all} $T_i \in \pmb{T}$}\label{alg2:start-tasks}
      \State{$satisfiedList \leftarrow \{\}$}
      \State{$unsatisfiedList \leftarrow \{\}$}
    \For {\textbf{all} $N_j \in \pmb{N}$}\label{alg2:start-nodes}
        \If{$T_i^{mem}\leq N_j^{mem}$}\label{alg2:iftask}
        \State{calculate $T_i^{response}$ using eq. \eqref{eq:eq6-1}} \label{alg2:iftask-response}
        \If{$T_i^{response} \leq T_i^{deadline}$}\label{alg2:iftask-response-deadline-start}
            \State $satisfiedtList\leftarrow satisfiedtList \cup \{j\}$
        \Else
            \State $unsatisfiedtList\leftarrow unsatisfiedtList \cup \{j\}$
        \EndIf \label{alg2:iftask-response-deadline-stop}
       \EndIf
    \EndFor\label{alg2:stop-nodes}
   \If{$|satisfiedList|\geq 1$}\label{alg2:satisfiedlist-start}
   \State{call \textbf{minCompComm}\textbf{($T_i$, satisfiedList)}}
    \Else  \label{alg2:unsatisfiedlist-start}
    \State{call \textbf{minViol($T_i$,unsatisfiedtList)}}
    \EndIf \label{alg2:unsatisfiedlist-stop}
   \EndFor\label{alg2:stop-tasks}
   \EndProcedure
  \end{algorithmic}
 \end{algorithm}
\end{center}

 The pseudo-code of Min-V is provided in Algorithm~\ref{alg:Min-V}. The detail of the algorithm is as follows. First of all, the available time of all nodes is set to 0 (line 2). In line~\ref{alg2:ascsort}, the algorithm sorts the tasks based on their deadlines in ascending order. Next, it iterates through tasks to allocate them to the most suitable nodes (lines~\ref{alg2:start-tasks} to~\ref{alg2:stop-tasks}). For each task, Min-V creates two empty lists: satisfiedList and unsatisfiedList (lines 5 and 6). In lines \ref{alg2:start-nodes} to \ref{alg2:stop-nodes}, for all nodes $N_j \in \pmb{N}$ which meet the memory demand of a given task $T_i$, the response time of task $T_i$ on node $N_j$ is obtained. If the node can satisfy the deadline of the task, it is added to satisfiedList; otherwise, it is added to unsatisfiedList. After this process, regarding the size of satisfiedList, two different strategies will be available. If there exists at least one node in satisfiedList, i.e., the deadline requirement of task $T_i$ is met, the algorithm calls \textit{minCompComm function} (see Algorithm~\ref{func:minCompComm}); otherwise it calls \textit{minViol function} (see Algorithm~\ref{func:minViol}).

 Algorithm~\ref{func:minCompComm} shows the pseudo-code of the minCompComm function. First of all it sets $C^{min} \leftarrow \infty$ in line 2. Then, the node with the minimum computation and communication cost is selected
(lines~\ref{func1:start-nodes-in-satisfiedList} to \ref{func1:stop-nodes-in-satisfiedList} ) and the task is assigned to the node and its available time is updated (lines 12 and 13).The pseudo-code of the minViol function is given in Algorithm~\ref{func:minViol}. Here again, we set $C^{min} \leftarrow \infty$ in line 2. Next, the node with the least violation cost is found (lines \ref{func2:start-nodes-in-unsatisfiedList} to \ref{func2:stop-nodes-in-unsatisfiedList}), and allocation and updating is done (lines 10 and 11).

\begin{center}
\alglanguage{pseudocode}
 \begin{algorithm}
  \allowdisplaybreaks
  \caption{minCompComm function}
  \label{func:minCompComm}
  \begin{algorithmic}[1]
   \Procedure{minCompComm}{$T_i$, satisfiedList}
   \State{$C^{min} \leftarrow \infty$}
    \For {\textbf{all} $N_j \in satisfiedList$}\label{func1:start-nodes-in-satisfiedList}
        \State {calculate $C_i^{comp}$ using eq.~\eqref{eq:eq3}}
    \State {calculate $C_i^{comm}$ using eq.~\eqref{eq:eq5}}
    \State{$C^{tot} \leftarrow C_i^{comp}+C_i^{comm}$}
    \If{$C^{tot} < C^{min}$}
    \State {$C^{min} \leftarrow C^{tot}$}
    \State {$index \leftarrow j$}
   \EndIf \label{func1:satisfiedlist-stop}
    \EndFor \label{func1:stop-nodes-in-satisfiedList}
        \State {allocate $T_i$ to $N_{index}$}
    \State{$availableTime [N_{index}] \leftarrow availableTime [N_{index}] + E_{i,index}$}
   \EndProcedure
  \end{algorithmic}
 \end{algorithm}
\end{center}

\begin{center}
\alglanguage{pseudocode}
 \begin{algorithm}
  \allowdisplaybreaks
  \caption{minViol function}
  \label{func:minViol}
  \begin{algorithmic}[1]
   \Procedure{minViol}{$T_i$, unsatisfiedtList}
   \State{$C^{min} \leftarrow \infty$}
      \For {\textbf{all} $N_j \in unsatisfiedList$}\label{func2:start-nodes-in-unsatisfiedList}
     \State {calculate $C_i^{viol}$ using eq.~\eqref{eq:eq7}}
         \If{$C_i^{viol} < C^{min}$}
            \State {$C^{min} \leftarrow C_i^{viol}$}
            \State {$index \leftarrow j$}
          \EndIf
         \EndFor \label{func2:stop-nodes-in-unsatisfiedList}
     \State {allocate $T_i$ to $N_{index}$}
    \State{$availableTime [N_{index}] \leftarrow availableTime [N_{index}] + E_{i,index}$}
   \EndProcedure
  \end{algorithmic}
 \end{algorithm}
\end{center}

\subsection{Illustrative Example}\label{proposedmethod:example}

\begin{figure}
	\centering
	\begin{subfigure}{0.48\textwidth}
		\centering
		\includegraphics[width=\linewidth]{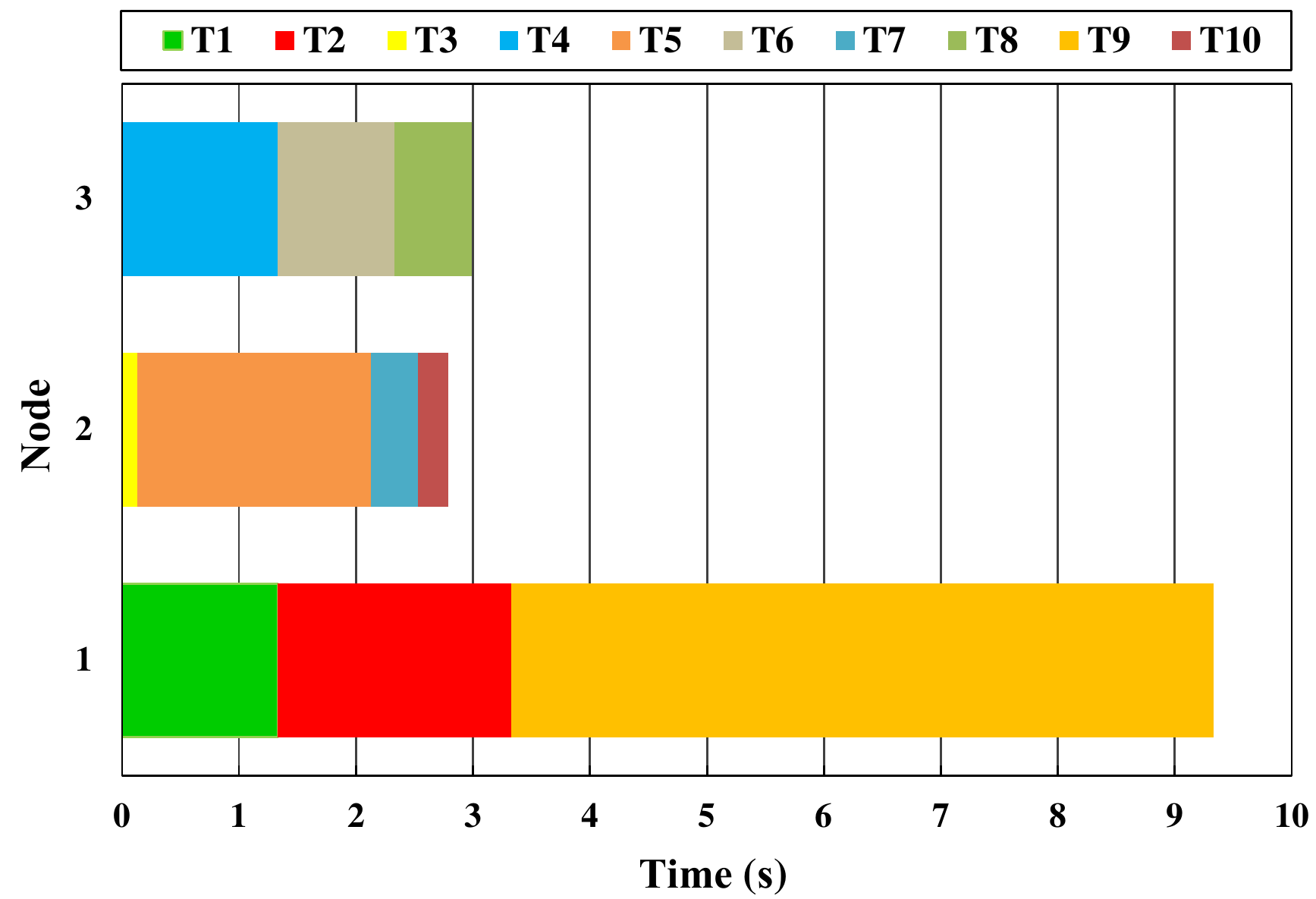}
		\caption{Min-CCV}
		\label{fig:illus-alg1}
	\end{subfigure}
	\begin{subfigure}{0.48\textwidth}
		\centering
		\includegraphics[width=\linewidth]{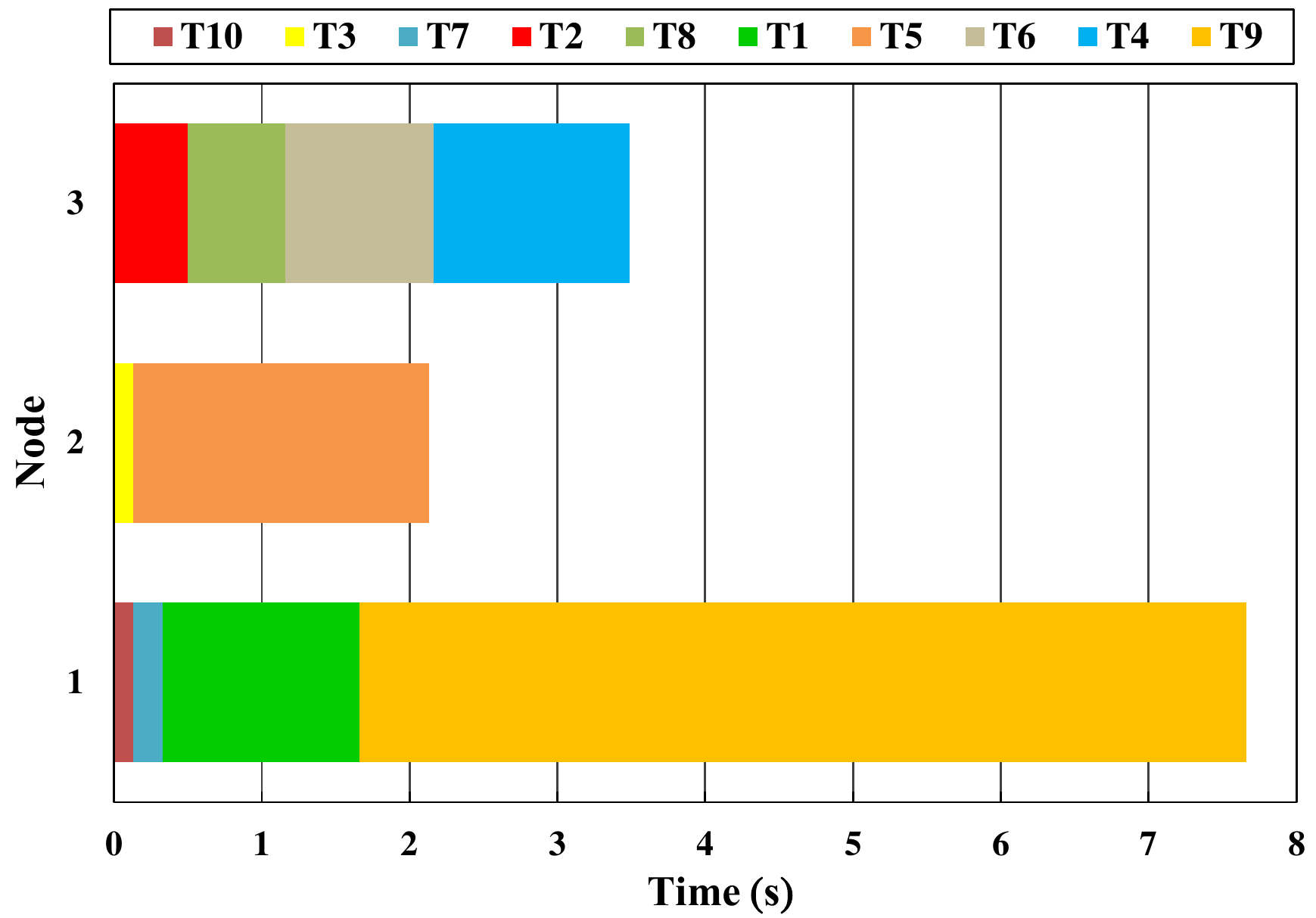}
		\caption{Min-V}
		\label{fig:illus-alg2}
	\end{subfigure}
	\caption{Toy case study allocation map.}
	\label{fig:fig3}
\end{figure}

In this part, we assume an illustrative example with 10 tasks and 3 nodes consist of 2 fog nodes and 1 cloud node. The tasks are numbered from 1 to 10 and nodes from 1 to 3. The attribute of tasks and nodes are shown in Table~\ref{tab:tab3} and Table \ref{tab:tab4}, respectively. Fig.~\ref{fig:fig3}a and Fig.\ref{fig:fig3}b respectively demonstrate the allocation map obtained by Min-CCV and Min-V. In Min-CCV, five of ten tasks cannot meet their deadlines where the total violation cost is about 505 $[G\$]$. However, in Min-V, only three tasks exceed their deadline while the total violation cost is 8.4 $[G\$]$.

\begin{table}
	\begin{center}
		\caption{Toy case study of attribute of tasks.}\label{tab:tab3}
		\footnotesize
		\begin{adjustbox}{max width=\textwidth}
	\begin{tabular}{|p{3.5cm}|c|c|c|c|c|c|c|c|c|c|}
				\hline
				 \multirow{2}{*}{\textbf{Parameters}}&  \multicolumn{10}{|c|}{\textbf{Tasks}}\\
				\cline{2-11}
	&\textbf{1}&\textbf{2}&\textbf{3}&\textbf{4}&\textbf{5}&\textbf{6}&\textbf{7}&\textbf{8}&\textbf{9}&\textbf{10}\\\hline
	\textbf{Number of instruction [MI]}&2000&3000&100&8000&1500&6000&300&4000&9000&200\\\hline
	\textbf{Memory required [MB]}&100&200&50&180&70&120&150&180&100&150\\\hline
	\textbf{Input file size [MB]}&0.5&1&0.3&1.5&0.4&1.2&0.8&1&0.5&1.4\\\hline
	\textbf{Output file size [MB]}&0.1&0.8&0.5&0.5&0.8&1&0.5&0.6&0.4&0.2\\\hline
	\textbf{Deadline [ms]}&1500&1000&200&5000&2200&3500&400&1200&8000&100\\\hline
	\textbf{QoS [\%]}&96&93&95&92&99&94&98&91&90&95\\\hline
	\textbf{Penalty [per \%]}&0.2&0.1&0.4&0.5&0.2&0.1&0.3&0.3&0.4&0.1\\\hline
			\end{tabular}
		\end{adjustbox}
	\end{center}
\end{table}

\begin{table}
	\begin{center}
		\caption{Toy case study of attribute of nodes.}\label{tab:tab4}
		\footnotesize
		\begin{adjustbox}{max width=\textwidth}
	\begin{tabular}{|p{4cm}|c|c|c|}
				\hline
				 \multirow{2}{*}{\textbf{Parameters}}&  \multicolumn{3}{|c|}{\textbf{Nodes}}\\
				\cline{2-4}
	&\textbf{1}&\textbf{2}&\textbf{3}\\\hline
	\textbf{CPU processing rate [MIPS]}&1500&750&6000\\\hline
	\textbf{CPU usage cost [G\$/s]}&0.3&0.4&1.5\\\hline
    \textbf{Memory usage cost [G\$/MB]}&0.03&0.02&0.05\\\hline
    \textbf{Bandwidth usage cost [G\$/MB]}&0.01&0.02&0.08\\\hline
    \textbf{Memory [MB]}&220&170&1024\\\hline
    \textbf{Delay [ms]}&1&2&150\\\hline
			\end{tabular}
		\end{adjustbox}
	\end{center}
\end{table}

\subsection{Complexity Analysis}\label{proposedmethod:complexity}
In this subsection, we provide the time and space complexity analysis of the our heuristic methods. 

\noindent 1) \textbf{Min-CCV:} The time complexity analysis of the Min-CCV algorithm is quite simple, where it depends on the number of tasks and nodes. To set the available time of nodes to 0 (Line 2), the algorithm calculates it in the order of $O(m)$. The rest of the algorithm consists of two nested loops where they run $n$ and $m$ times, respectively. Therefore, the overall time complexity of this algorithm is $O(n \times m)$.  Moreover, the space complexity of Min-CCV is linear in the number of tasks and nodes, i.e., $O(n+m)$.  

\noindent 2) \textbf{Min-V:} Similar to Min-CCV, the complexity of Line 2 is $O(m)$. According to the line \#3, Min-V sorts the set of $n$ tasks, which requires $O(nlogn)$. To construct \emph{satisfiedList} and \emph{unsatisfiedList}, the algorithm runs two nested \textbf{for} loops which calculates in the order of $O(n \times m)$ time. The rest of the algorithm depends on the size of \emph{satisfiedList}. If this list is not empty, it may include all of the nodes; thus searching among them needs $O(m)$ time in the worst case. Otherwise, the algorithm searches among the nodes inside the \emph{unsatisfiedList} that runs in the order of $O(m)$. Hence, the overall time complexity of the Min-V algorithm is $O(nlogn+n \times m)$.  The space complexity of this algorithm is also linear in the number of tasks and nodes, i.e., $O(n+m)$. 

The above analysis shows that our proposed algorithms can run very fast, which make them a promising solution for real-time volunteer task scheduling problem in fog-cloud computing systems.  

\section{Performance Evaluation}\label{simulation}
In this section, we describe simulation setup, performance metrics and the results. Our proposed algorithms are compared with three cutting-edge strategies, and the results verify that our proposed algorithms are superior to other algorithms.

\subsection{Simulation setup}\label{simulation-setup}

Here, we describe the simulation settings (see Section~\ref{setting}), simulation metrics (see Section~\ref{metrics}), comparison algorithms (see Section~\ref{comparing}), and the results (see Section~\ref{results}).

\subsubsection{Settings}\label{setting}
In our experimental study, to fully apprehend the benefits of the proposed algorithms, we have performed three experiments in which we investigate the impact of the various parameters that determine the results. The purpose of the experiments, along with their settings, are outlined in Table~\ref{tab: experiment-settings}. They are also briefly explained here.

In this paper, we consider three various experiments as below
\begin{itemize}[leftmargin=*]
    \item \textbf{Experiment one}: In this experiment, in each level, we increase 50 tasks to the number of tasks from 50 to 300 tasks, and the number of fog and cloud nodes fixed to 30 and 15 respectively (see Fig.~\ref{fig:fig4}).
    \item \textbf{Experiment two}: In this experiment, the problem consists of 200 tasks and 15 cloud nodes with a varying number of fog nodes from 10 to 50 (see Fig.~\ref{fig:fig5}).
     \item \textbf{Experiment three}: In this experiment, we want to observe the impact of the varying number of cloud nodes, with the number of tasks and fog nodes fixed to 200 and 30, respectively (see Fig.~\ref{fig:fig6}).
\end{itemize}

\begin{table}
\caption{Simulation experiment settings}\label{tab: experiment-settings}
\footnotesize
  \begin{tabular}{|c|c|c|c|c|}
  \hline
  \multirow{2}{*}{\textbf{Experiment}}&\multirow{2}{*}{\textbf{Purpose of Experiment}}&\multicolumn{3}{|c|}{\textbf{Parameters}}\\\cline{3-5}
  &&\textbf{Task} &\textbf{Fog Node}&\textbf{Cloud Node}\\\hline
  1&Impact of varying number of tasks&[50, 300]&30&15\\\hline
  2&Impact of varying number of fog nodes&200&[10, 50]&15\\\hline
  3&Impact of varying number of cloud nodes&200&30&[5, 25]\\\hline
  \end{tabular}
  \end{table}

The fog-cloud system is responsible for executing all IoT requests. Each request first is decomposed into a set of independent tasks; then their resources requirement are calculated and estimated. Table~\ref{tab: experiment-task-settings} shows the characteristics of tasks used in our experiments.  We run each experiment ten times and report the average, maximum and minimum of them.

\begin{table}
\caption{Attribute of task settings}\label{tab: experiment-task-settings}
\footnotesize
  \begin{tabular}{|c|c|c|c|c|}
  \hline
  \multirow{2}{*}{\textbf{Parameter}}&\multicolumn{3}{|c|}{\textbf{Value}}&\multirow{2}{*}{\textbf{Unit}}\\\cline{2-4}
  &\textbf{Type 1} &\textbf{Type 2}&\textbf{Type 3}&   \\\hline
  \textbf{Size}&[100, 372]&[1028, 4280]&[5123, 9784]& [MI]\\\hline
  \textbf{Required memory}&\multicolumn{3}{|c|}{[50,200]}& [MB]\\\hline
  \textbf{Input file size}&\multicolumn{3}{|c|}{[0.3,1.5]}& [MB]\\\hline
  \textbf{Output file size}&\multicolumn{3}{|c|}{[0.1,1]}& [MB]\\\hline
  \textbf{Deadline}&[100, 500]&[500, 2500]&[2500, 10000]& [ms]\\\hline
  \textbf{QoS}&\multicolumn{3}{|c|}{[90,99.99]}& [\%]\\\hline
  \textbf{Penalty}&\multicolumn{3}{|c|}{[0.1,0.5]}& [G\$/\%]\\\hline
  \end{tabular}
  \end{table}

The fog-cloud infrastructure consists of fog nodes with limited process capacity than the cloud nodes, but they are closer to the user and have a minimum delay. Consequently, cloud nodes can process the tasks in the shortest possible time; however, they have a high delay in receiving tasks. Hence, the proposed algorithms should be handling the balancing between fog and cloud nodes to decrease the total cost. Table~\ref{tab: experiment-node-settings} presents the attributes of nodes with the process capacity details and the minimum delay in allocating tasks on nodes. All of the attribute values of the nodes can be chosen randomly. We perform experiments 2 and 3 with 10 iterations to investigate the impact of the number of nodes.
  \begin{table}
\caption{Attribute of node settings}\label{tab: experiment-node-settings}
\footnotesize
  \begin{tabular}{|c|c|c|c|c|}
  \hline
  \multirow{2}{*}{\textbf{Parameter}}&\multicolumn{2}{|c|}{\textbf{Value}}&\multirow{2}{*}{\textbf{Unit}}\\\cline{2-3}
  &\textbf{Fog} &\textbf{Cloud}&   \\\hline
  \textbf{CPU processing rate}&[500, 2000]&[3000, 10000]& [MIPS]\\\hline
  \textbf{CPU usage cost}&[0.2,0.5]&[1, 2.1]& [G\$/second]\\\hline
  \textbf{Memory usage cost}&[0.01,0.03]&[0.02,0.05]& [G\$/MB]\\\hline
  \textbf{Bandwidth usage cost}&[0.01,0.02]&[0.05,0.1]& [G\$/MB]\\\hline
   \textbf{Memory}&[150,250]&[256,4096]& [MB]\\\hline
   \textbf{Delay}&[1,5]&[50,250]& [ms]\\\hline
  \end{tabular}
  \end{table}
All experimental simulations were carried out on MATLAB environment on a PC with Intel(R) Xeon(R) CPU E7-4850 v4 @ 2.10 GHz (2 processors), 8.00 GB RAM, and Windows 10 pro operating system. The source code of our paper is available in~\cite{TOITFarooq2020}.

 \subsubsection{Simulation Metrics}\label{metrics}
 
In this paper, our purpose is to ensure the quality of service (QoS) with minimum total cost. To evaluate the performance of the proposed algorithms, we measured computation, communication, and violation cost. As the fog computing environment is suited explicitly to time-sensitive applications, violation cost is the most important factor which needs to be ensured. Therefore, we must try to find the best allocating map to decrease delays and makespan. On the other hand, the computation and communication costs should not be high.
 
  \subsubsection{Comparison Algorithms}\label{comparing}
  
Our proposed algorithms are compared against the three algorithms that consist of two base methods and a metaheuristic-based approach.
  
 \begin{itemize}[leftmargin=*]
  	\item \textbf{Round Robin (RR)}: in this method, the execution time for each task is estimated on the nodes one by one, and a node is chosen if the execution time lower than the deadline’s task and ensure the memory limitation. This operation will continue until any node has one task at least, then, the second round is begun.
  	\item \textbf{Random}: according to its name, the task is allocated on the nodes randomly with ensuring the memory limitation.
  	\item  \textbf{TCaS}: this is a metaheuristic algorithm introduced in \cite{8}. We consider 1000 generations in each iteration.
  \end{itemize}

\subsubsection{Results}\label{results}
In this part, we provide the results for the \textit{three} experiments mentioned above with two proposed algorithms and three comparing algorithms. In this paper, we calculated three costs for minimizing total cost. On the other hand, we used decreasing makespan and increasing the PDST (Percentage of the deadline satisfied tasks) for reducing the violation cost. Consequently, we present six plots for each experiment.

\begin{figure}
	\centering
	\begin{subfigure}{0.48\textwidth}
		\centering
		\includegraphics[width=\linewidth]{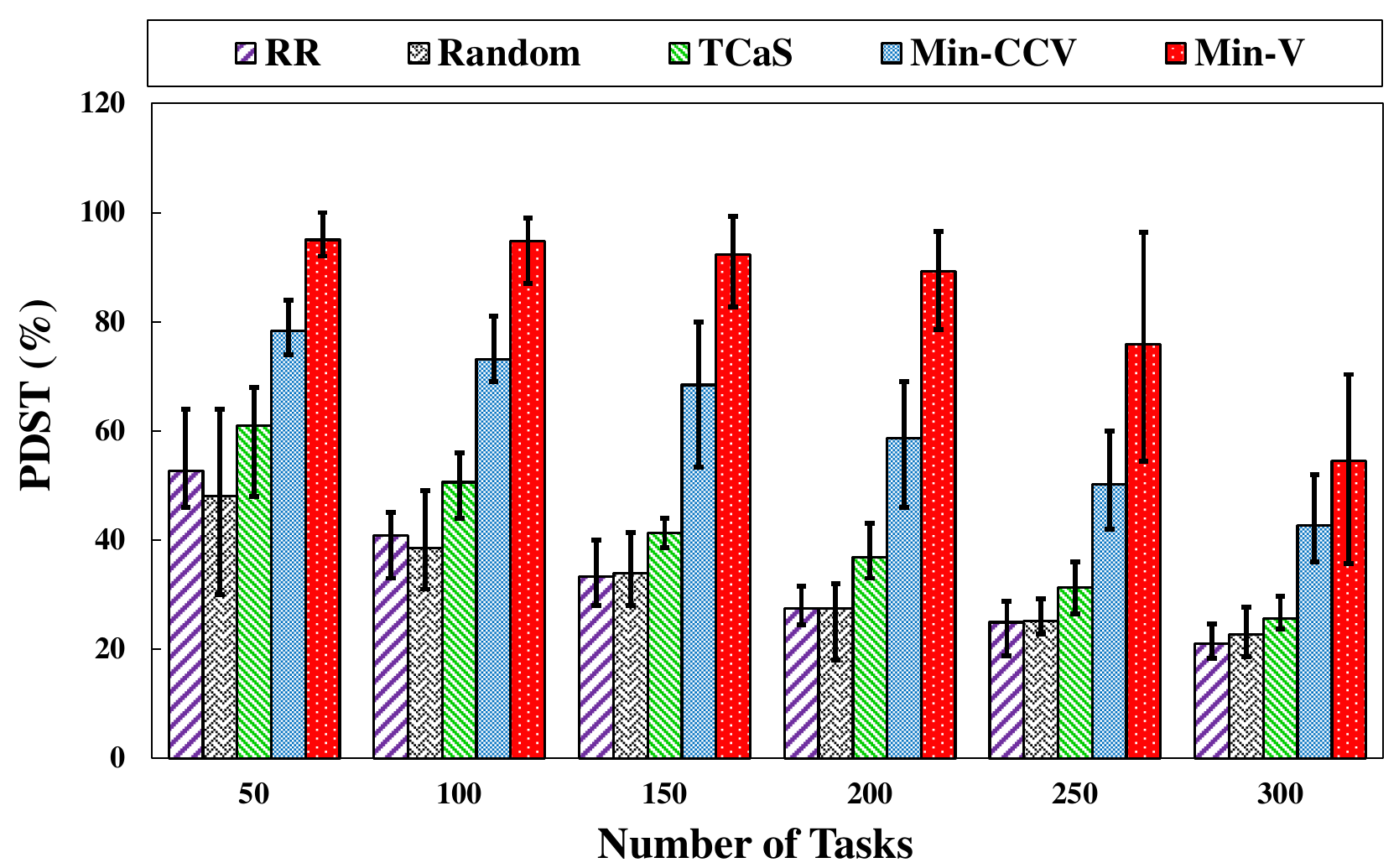}
		\caption{PDST}
		\label{fig:exp1-PDST}
	\end{subfigure}
	\begin{subfigure}{0.48\textwidth}
		\centering
		\includegraphics[width=\linewidth]{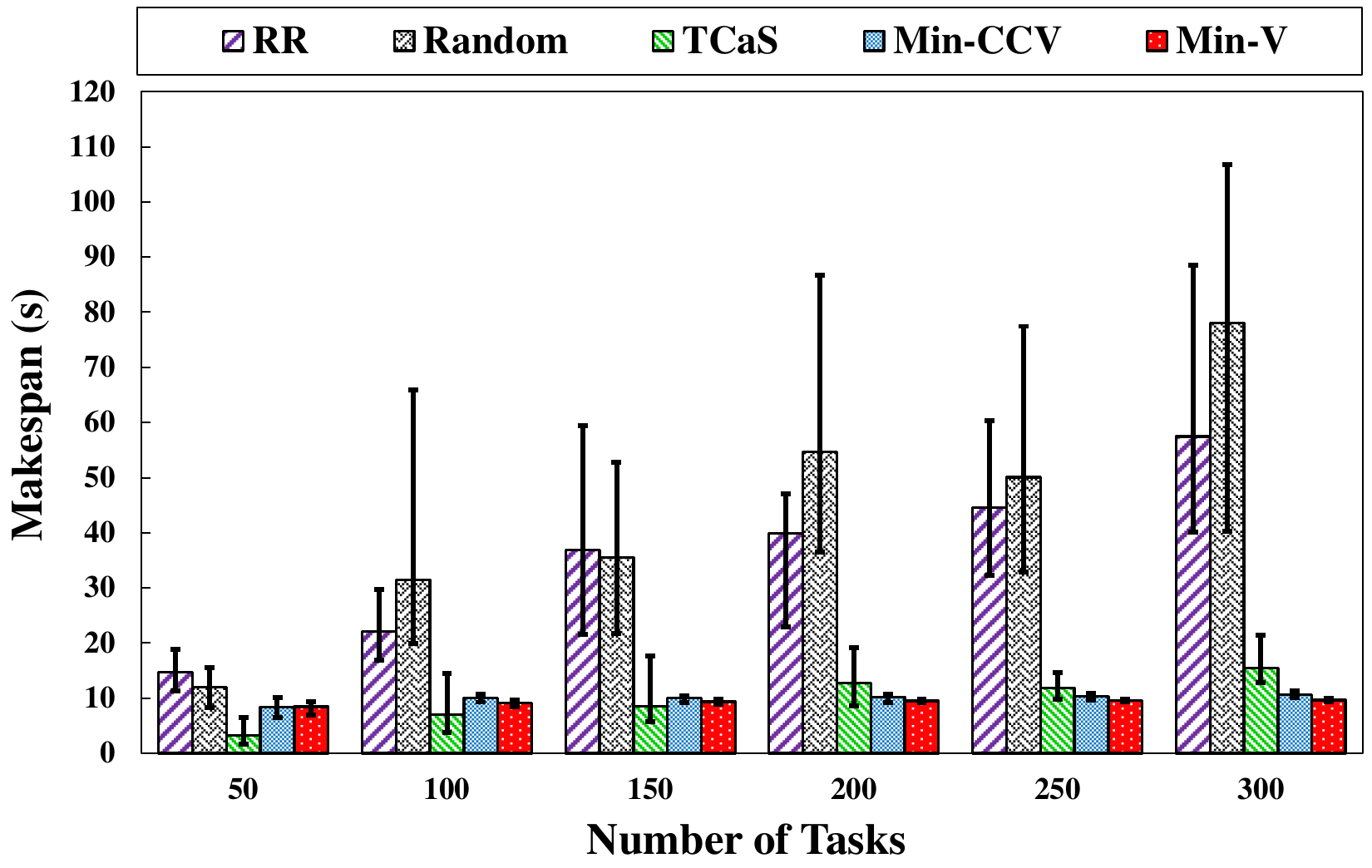}
		\caption{Makespan}
		\label{fig:exp1-Makespan}
	\end{subfigure}
	\begin{subfigure}{0.48\textwidth}
		\centering
		\includegraphics[width=\linewidth]{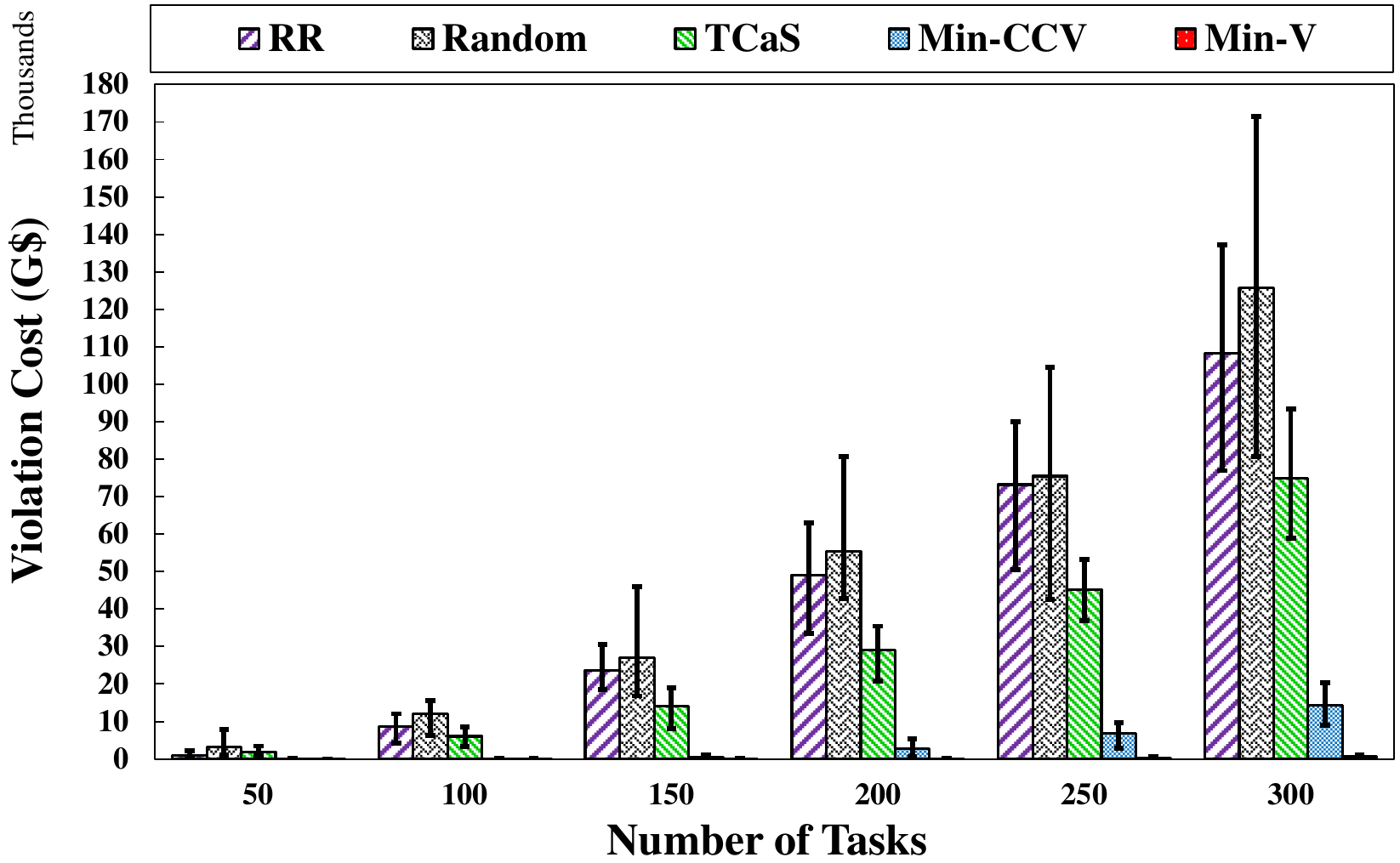}
		\caption{Violation cost}
		\label{fig:exp1-Violation}
	\end{subfigure}
	\begin{subfigure}{0.48\textwidth}
		\centering
		\includegraphics[width=\linewidth]{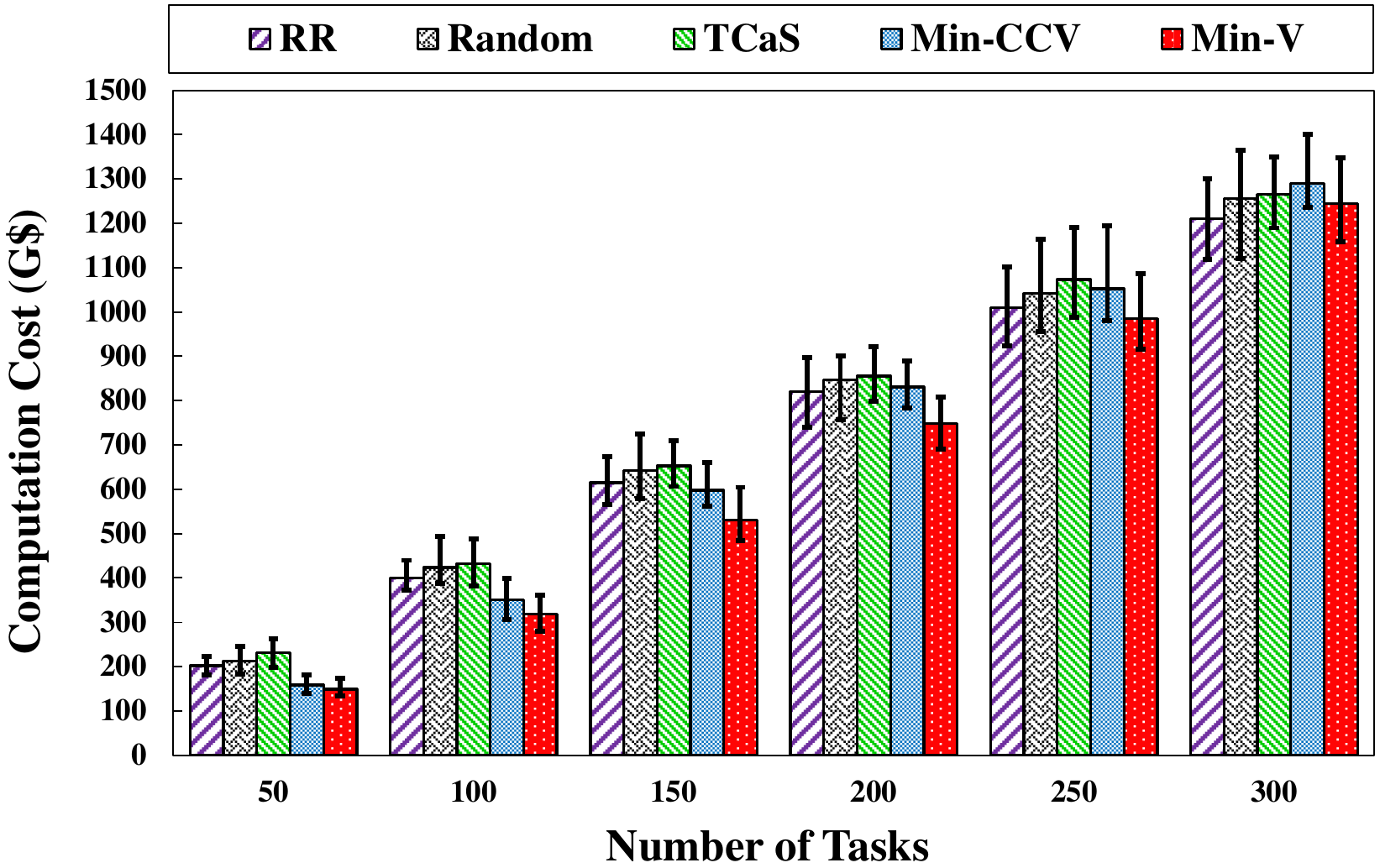}
		\caption{Computation cost}
		\label{fig:exp1-Computation}
	\end{subfigure}
	\begin{subfigure}{0.48\textwidth}
		\centering
		\includegraphics[width=\linewidth]{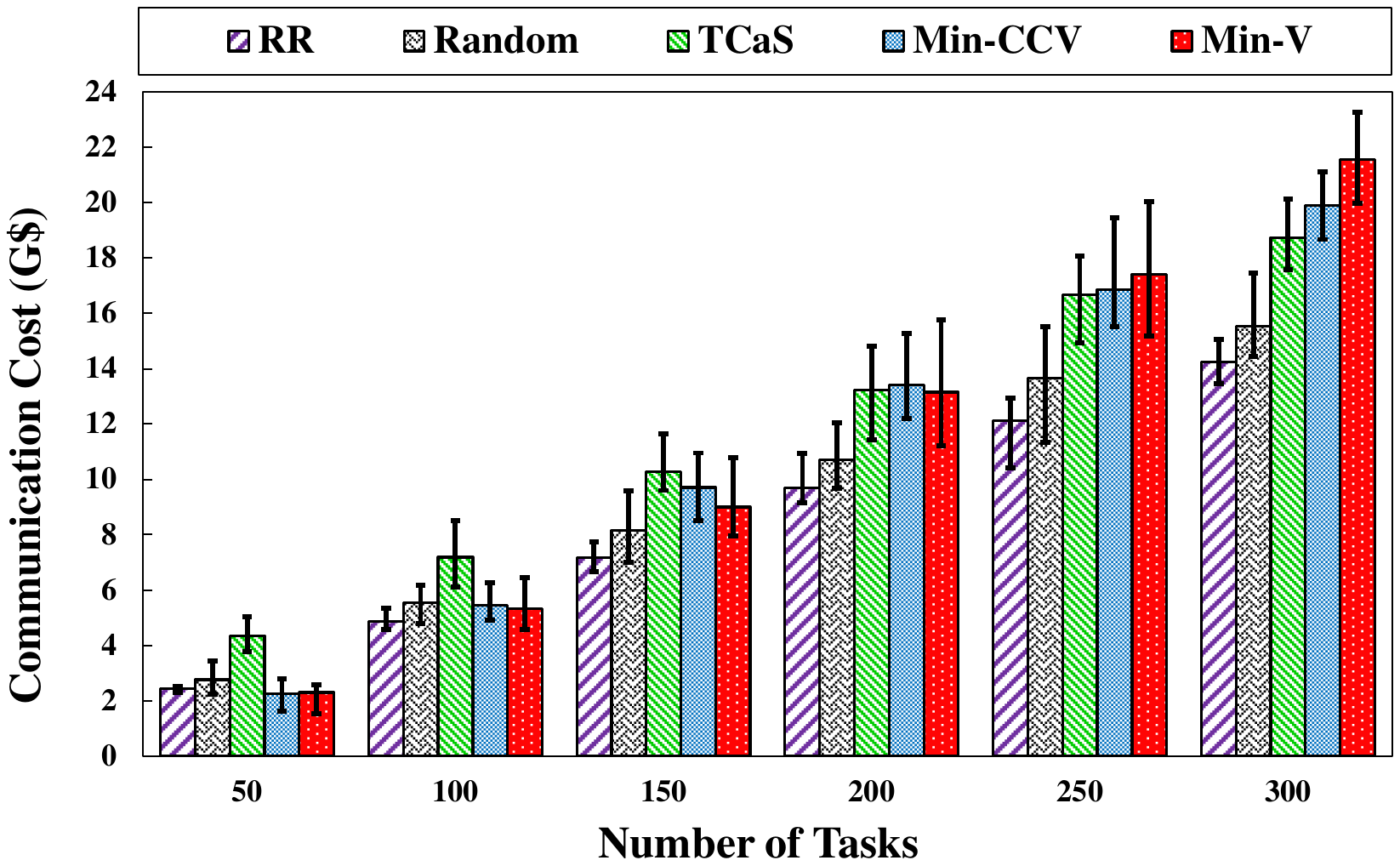}
		\caption{Communication cost}
		\label{fig:exp1-Communication}
	\end{subfigure}
	\begin{subfigure}{0.48\textwidth}
		\centering
		\includegraphics[width=\linewidth]{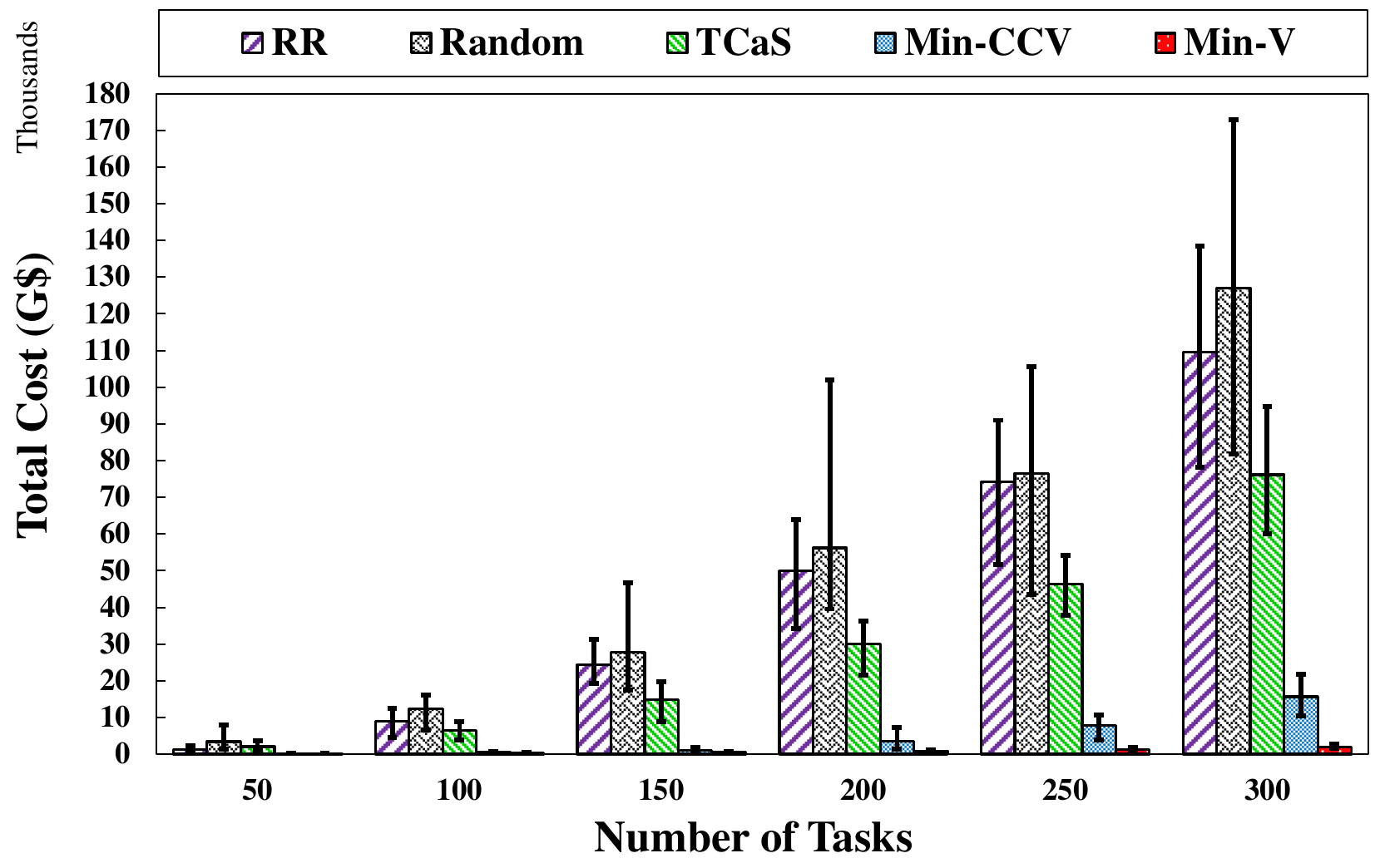}
		\caption{Total cost}
		\label{fig:exp1-Total}
	\end{subfigure}
	\caption{Simulation results for \textit{Experiment 1} (impact of varying number of tasks).}
	\label{fig:fig4}
\end{figure}

\textbf{1) Impact of a varying number of tasks:} Figs.~\ref{fig:fig4}a to ~\ref{fig:fig4}f illustrate the performance and cost of the proposed algorithms compared to the benchmarks. From the figures, we see that with increasing the number of tasks, the PDST decreases while the other metrics increases. However, our proposed algorithms present higher performance and lower cost compared with the others. Specifically, Min-V significantly outperforms the baseline methods and TCaS in terms of PDST (see Fig. \ref{fig:fig4}a) and violation cost (see Fig.\ref{fig:fig4}c). This is because of the deadline, violation-awareness of this algorithm. It also remarkably reduces makespan compared to RR and Random while it is a competitor for TCaS (see Fig.\ref{fig:fig4}b). The Min-CCV algorithm provides the second-best results in terms of PDST and the cost of the violation. Regarding the communication cost (Fig.~\ref{fig:fig4}e), since compared to the violation cost, the communication cost has less impact on the total cost, and this metric is somewhat higher for our algorithms. However, both of our algorithms perform well in terms of computation cost (Fig.~\ref{fig:fig4}d). Finally, Fig.~\ref{fig:fig4}f depicts that our Min-V and then Min-CCV provide extremely low total cost compared to the rest. 

\begin{figure}
	\centering
	\begin{subfigure}{0.48\textwidth}
		\centering
		\includegraphics[width=\linewidth]{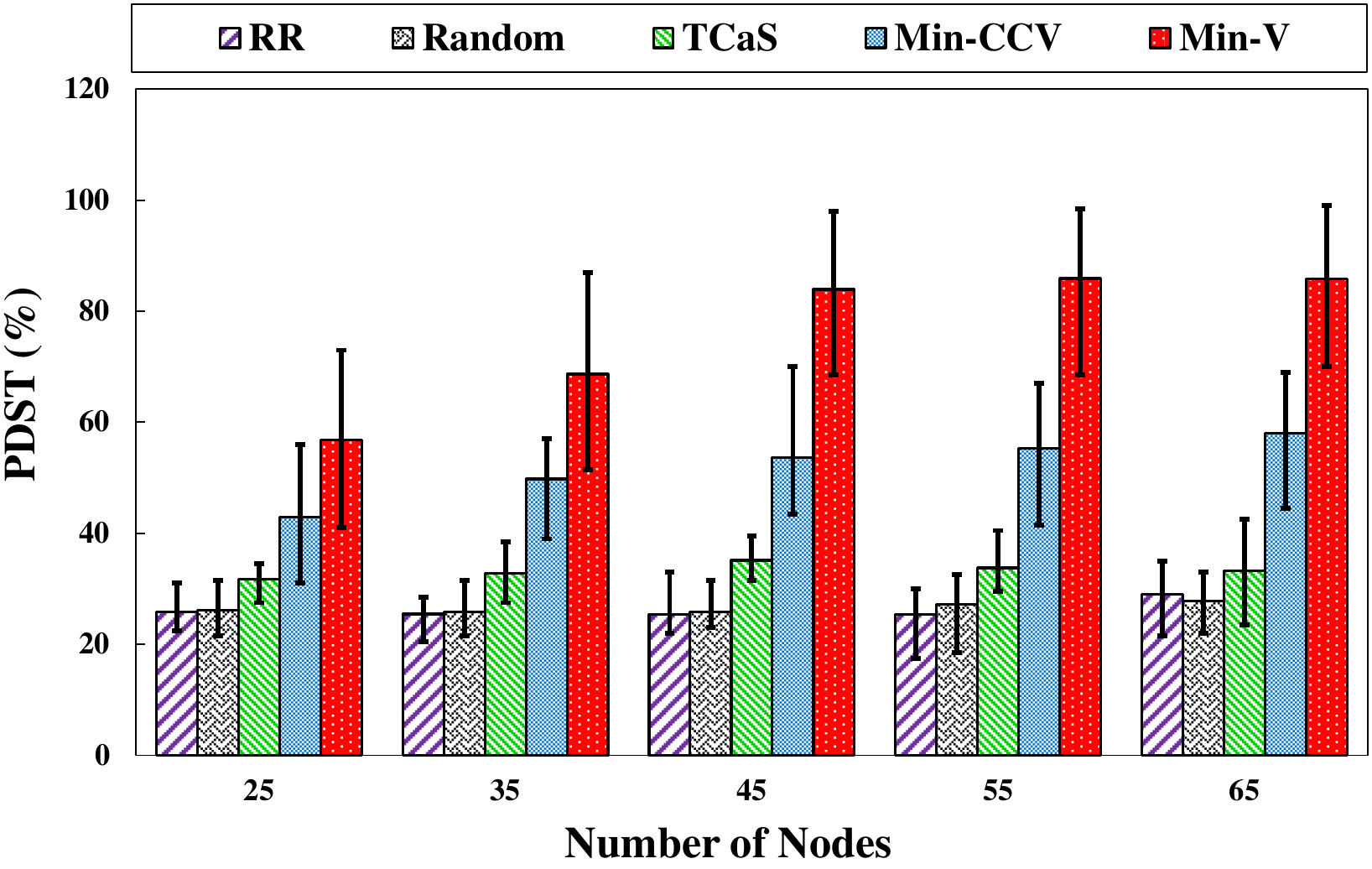}
		\caption{PDST}
		\label{fig:exp2-PDST}
	\end{subfigure}
	\begin{subfigure}{0.48\textwidth}
		\centering
		\includegraphics[width=\linewidth]{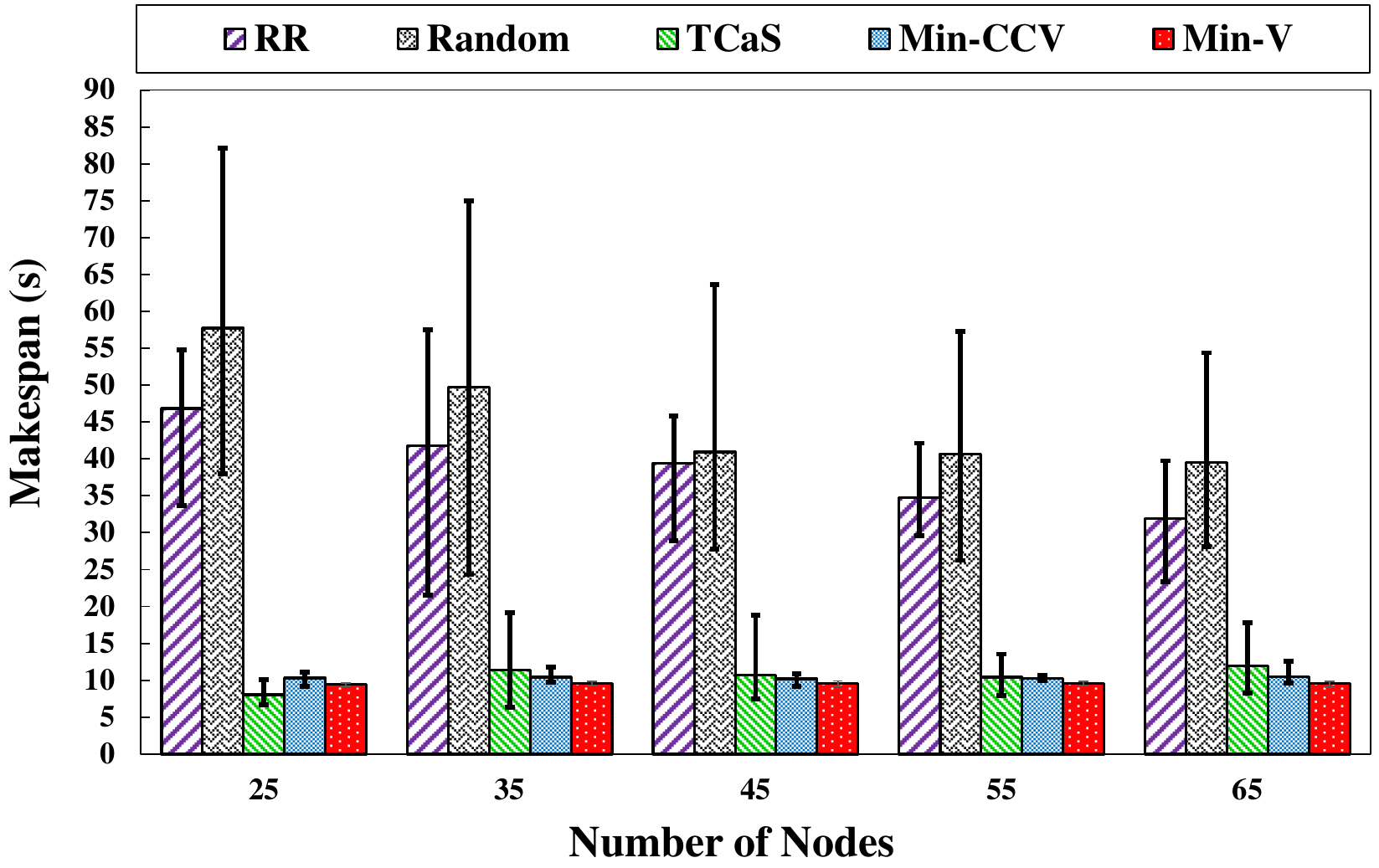}
		\caption{Makespan}
		\label{fig:exp2-Makespan}
	\end{subfigure}
	\begin{subfigure}{0.48\textwidth}
		\centering
		\includegraphics[width=\linewidth]{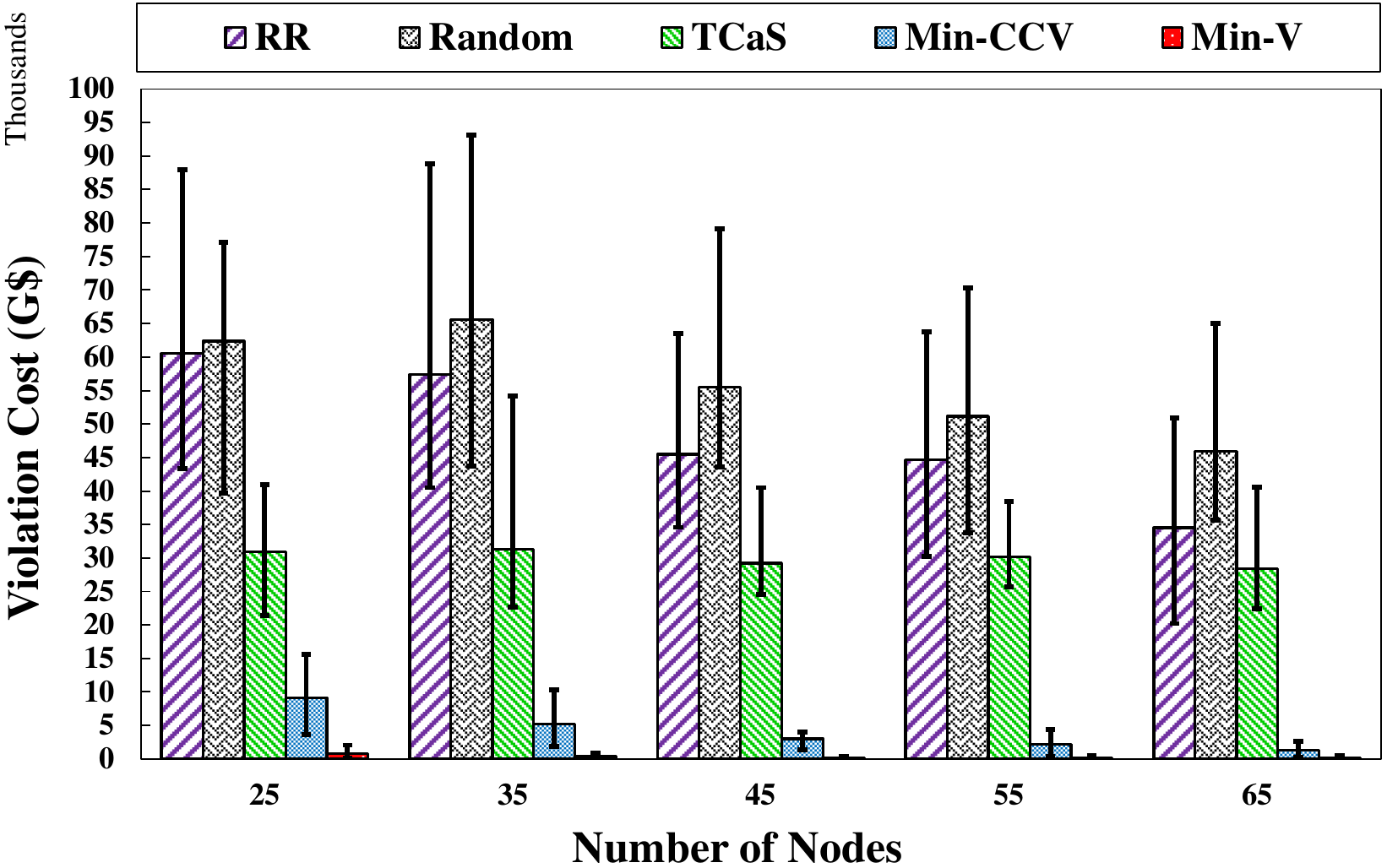}
		\caption{Violation cost}
		\label{fig:exp2-Violation}
	\end{subfigure}
	\begin{subfigure}{0.48\textwidth}
		\centering
		\includegraphics[width=\linewidth]{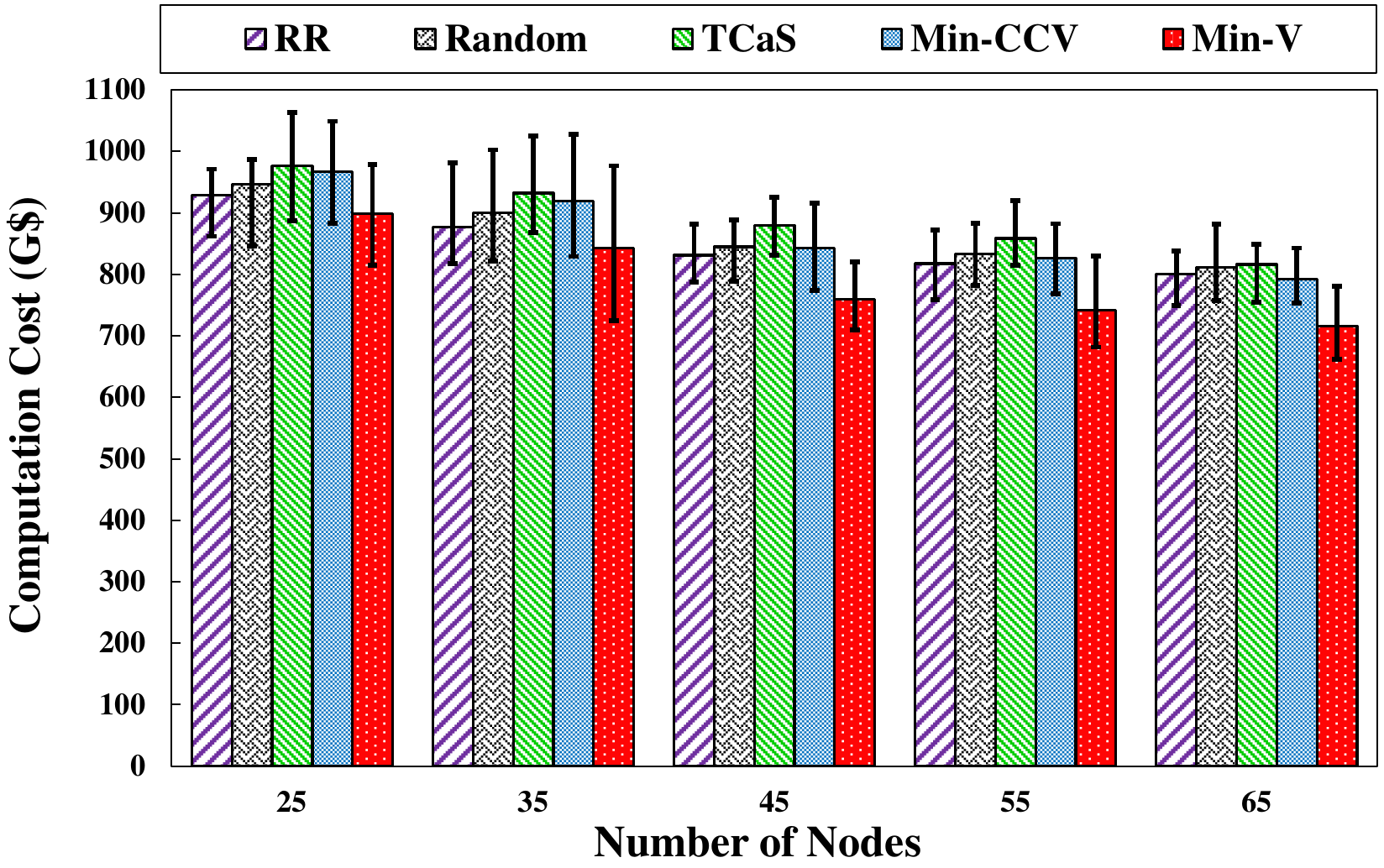}
		\caption{Computation cost}
		\label{fig:exp2-Computation}
	\end{subfigure}
	\begin{subfigure}{0.48\textwidth}
		\centering
		\includegraphics[width=\linewidth]{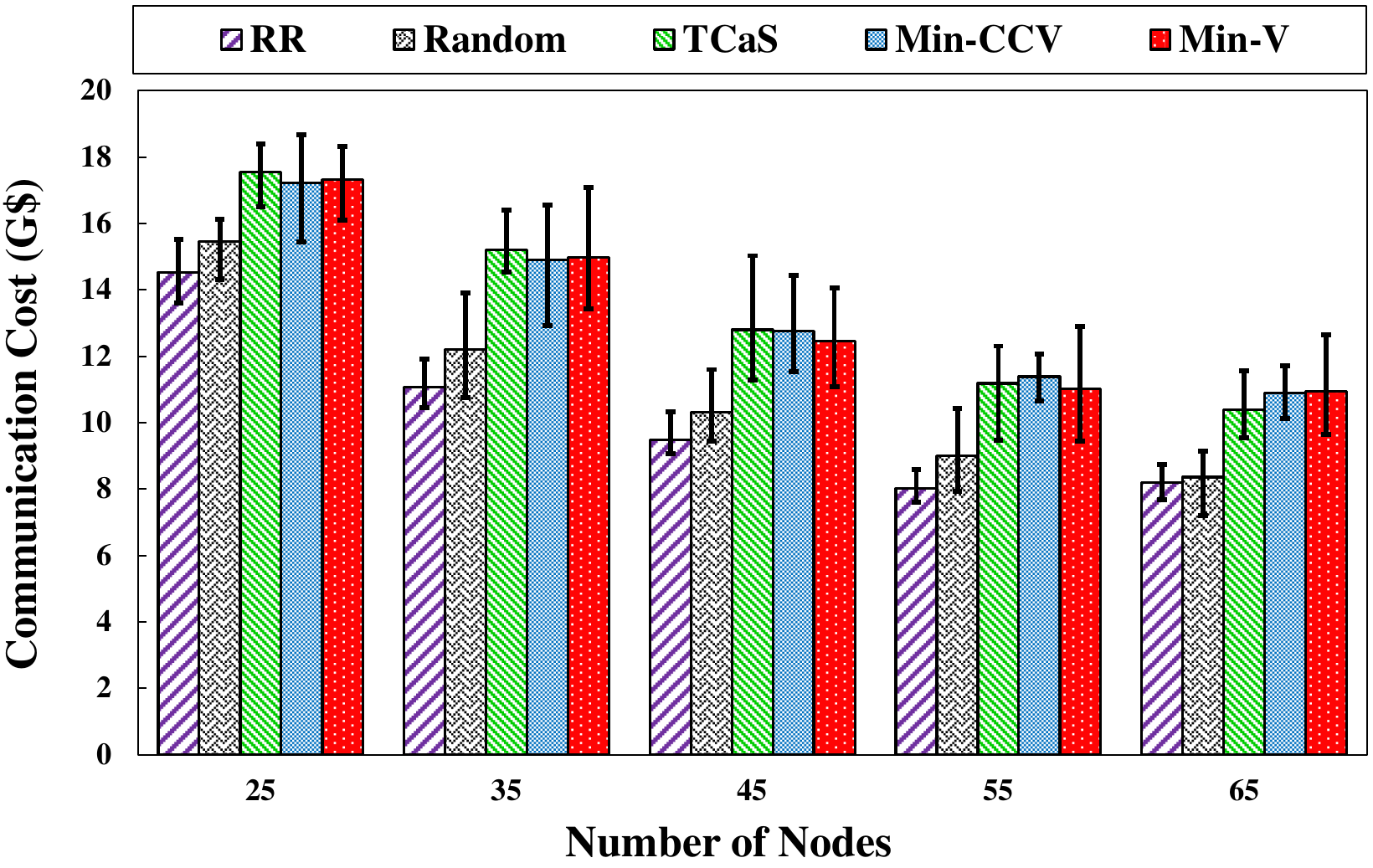}
		\caption{Communication cost}
		\label{fig:exp2-Communication}
	\end{subfigure}
	\begin{subfigure}{0.48\textwidth}
		\centering
		\includegraphics[width=\linewidth]{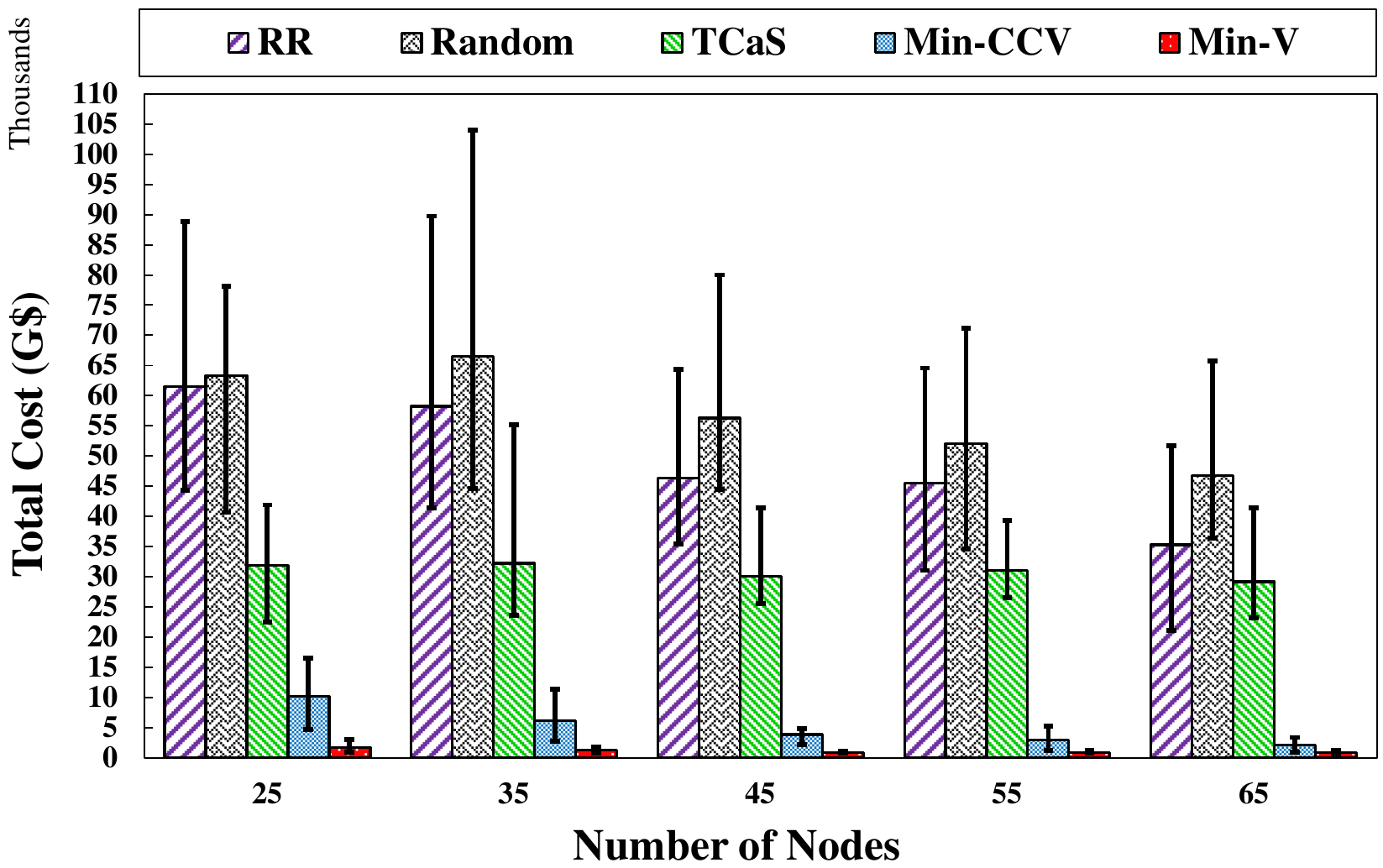}
		\caption{Total cost}
		\label{fig:exp2-Total}
	\end{subfigure}
	\caption{Simulation results for \textit{Experiment 2} (impact of varying number of fog nodes).}
	\label{fig:fig5}
\end{figure}

\textbf{2) Impact of varying number of fog nodes:} The results of this experiment are shown in Figs.~\ref{fig:exp2-PDST} to ~\ref{fig:exp2-Total}. Generally speaking, as the number of fog nodes increases, the PDST of all algorithms is increasing while the system cost decreases. However, due to prioritizing tasks based on their deadline requirements and taking into account the violation, computation and communication cost, the proposed Min-V gives the best performance compared to the rest. More surprisingly, as the number of fog nodes increases from 10 to 50, where the number of cloud nodes is fixed to 15, Min-V achieves 67 to 92\% of the PDST (Fig.~\ref{fig:exp2-PDST}) and reduces the violation cost from 574 to 85 [G$\$$] (Fig.~\ref{fig:exp2-Violation}). This implies that Min-V is capable of achieving great QoS for IoT requests. In these respects, Min-CCV provides the second-best results. 
As we can observe from Fig.~\ref{fig:exp2-Makespan}, in terms of makespan, our proposed algorithms and TCaS almost have the same performance while they are far better than RR and Random. From the point of computation and communication cost (Figs.~\ref{fig:exp2-Computation} and~\ref{fig:exp2-Communication}), as the number of fog nodes increases, all algorithms allocate more tasks to the fog environment which results in reducing the computation and communication cost. However, since the computation cost has a higher impact on the total cost, our proposed algorithms focus more on reducing it. Fig.~\ref{fig:exp2-Total} demonstrates the overall system cost. The percentage of improvement for our Min-V and Min-CCV is up to 99.5\% and about 90 to 98\% in compare with TCaS, respectively.

\begin{figure}
	\centering
	\begin{subfigure}{0.48\textwidth}
		\centering
		\includegraphics[width=\linewidth]{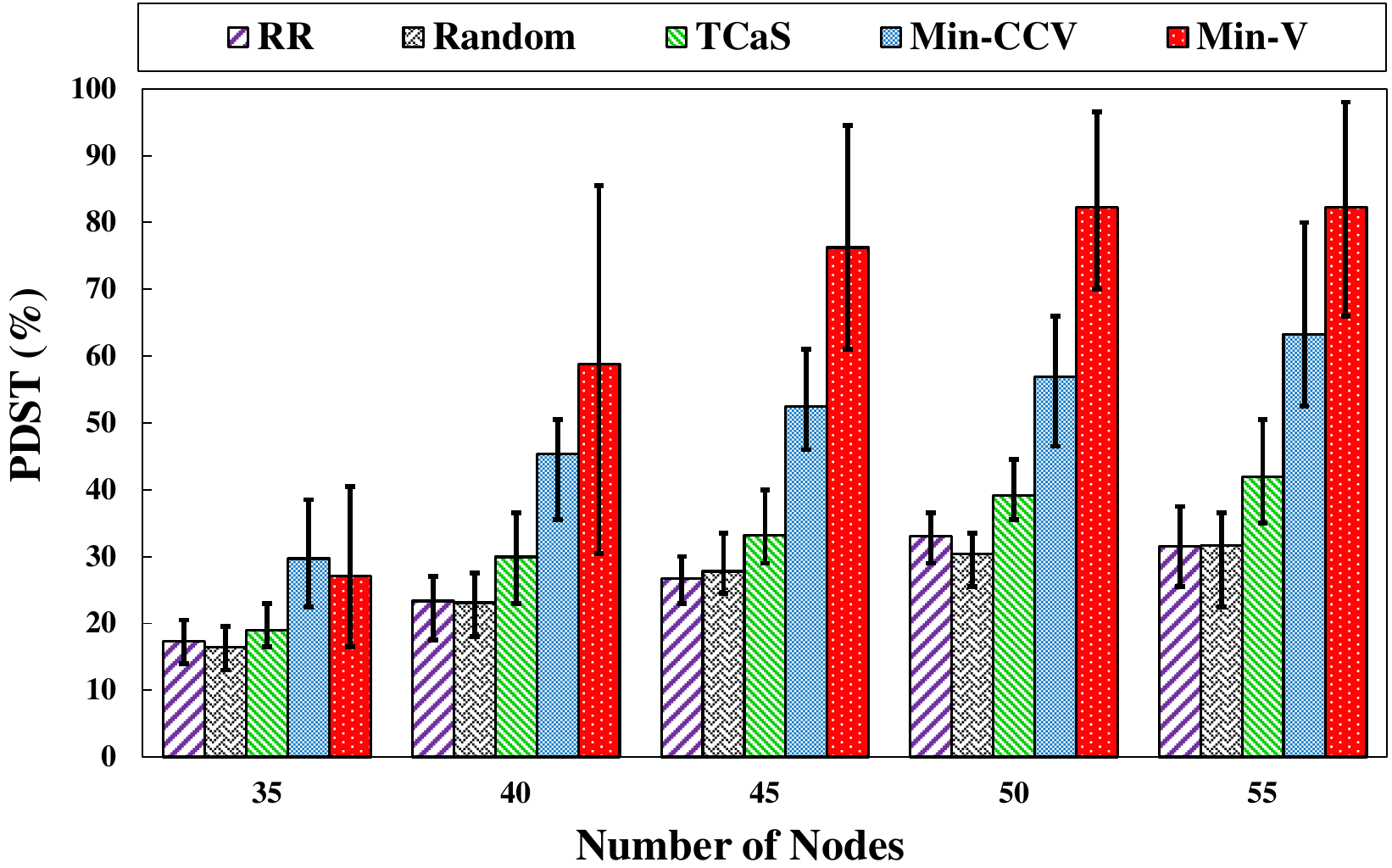}
		\caption{PDST}
		\label{fig:exp3-PDST}
	\end{subfigure}
	\begin{subfigure}{0.48\textwidth}
		\centering
		\includegraphics[width=\linewidth]{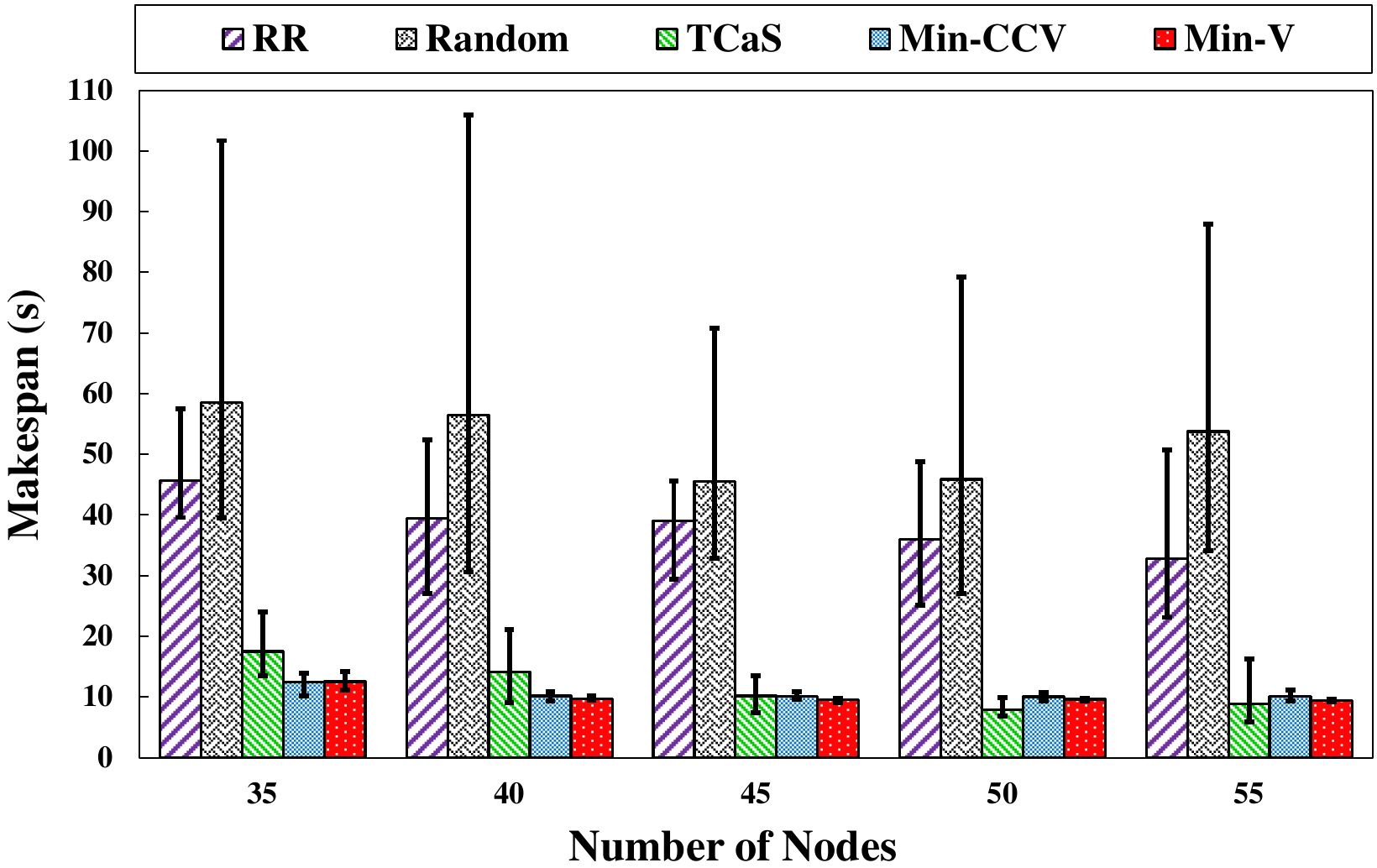}
		\caption{Makespan}
		\label{fig:exp3-Makespan}
	\end{subfigure}
	\begin{subfigure}{0.48\textwidth}
		\centering
		\includegraphics[width=\linewidth]{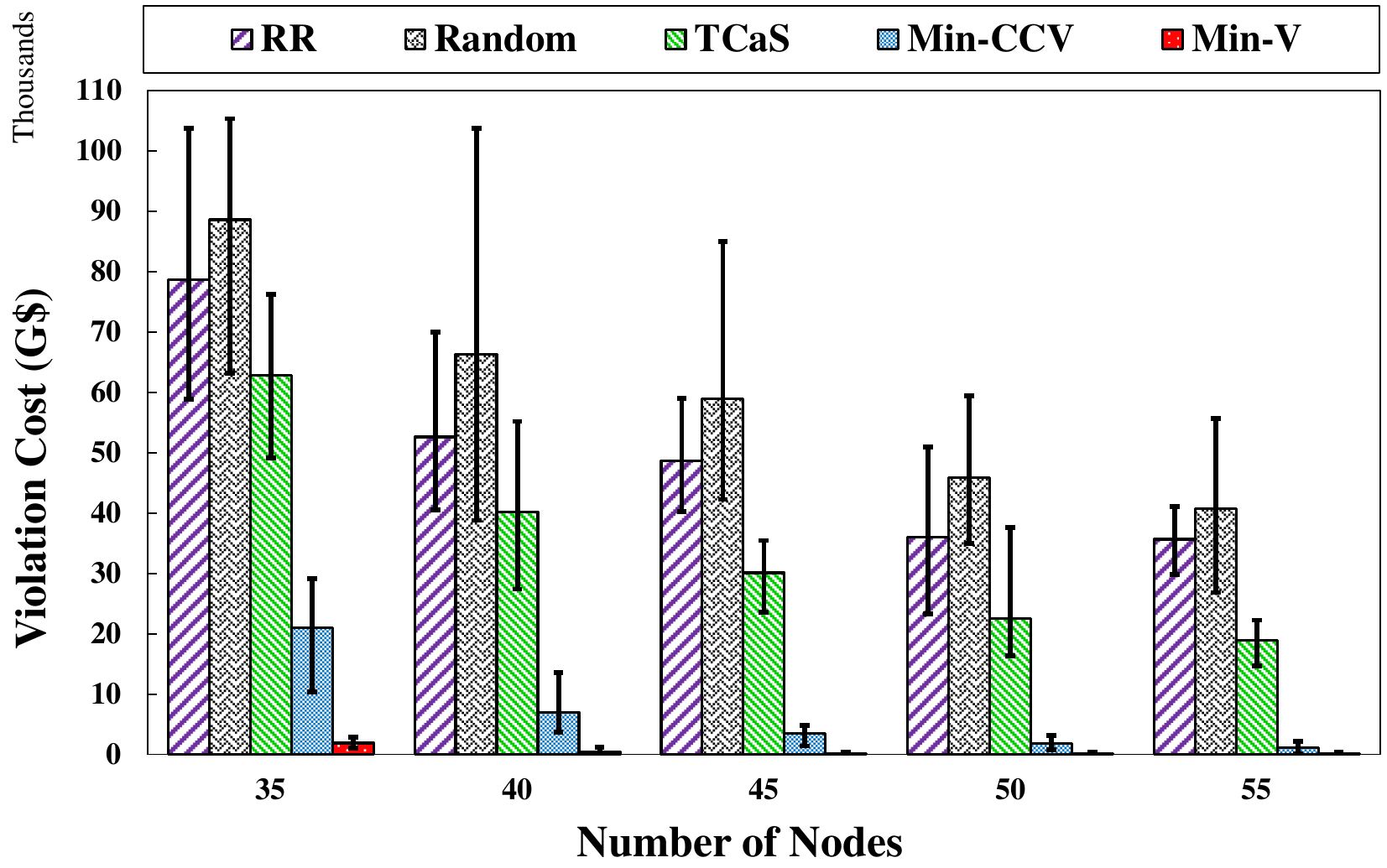}
		\caption{Violation cost}
		\label{fig:exp3-Violation}
	\end{subfigure}
	\begin{subfigure}{0.48\textwidth}
		\centering
		\includegraphics[width=\linewidth]{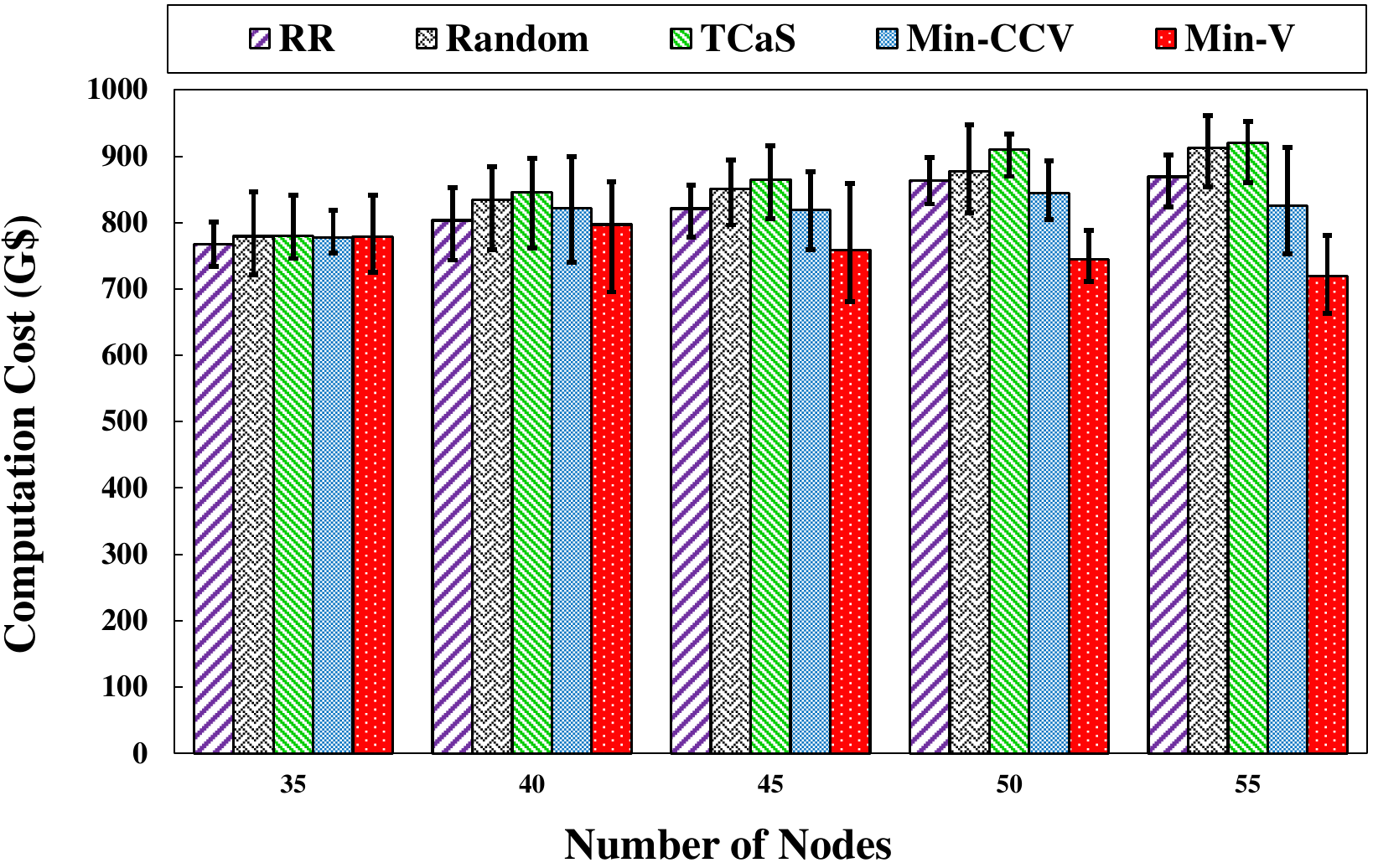}
		\caption{Computation cost}
		\label{fig:exp3-Computation}
	\end{subfigure}
	\begin{subfigure}{0.48\textwidth}
		\centering
		\includegraphics[width=\linewidth]{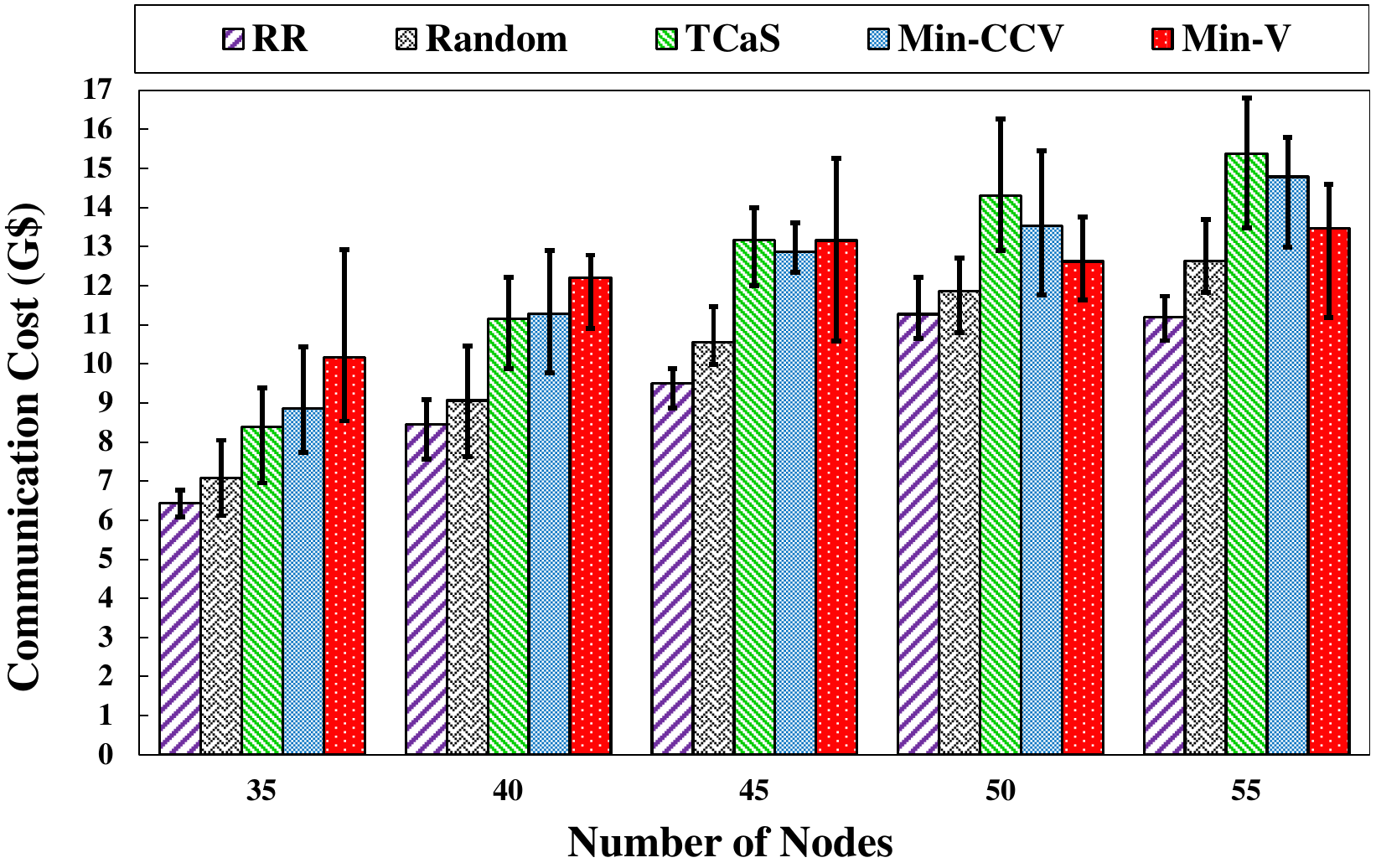}
		\caption{Communication cost}
		\label{fig:exp3-Communication}
	\end{subfigure}
	\begin{subfigure}{0.48\textwidth}
		\centering
		\includegraphics[width=\linewidth]{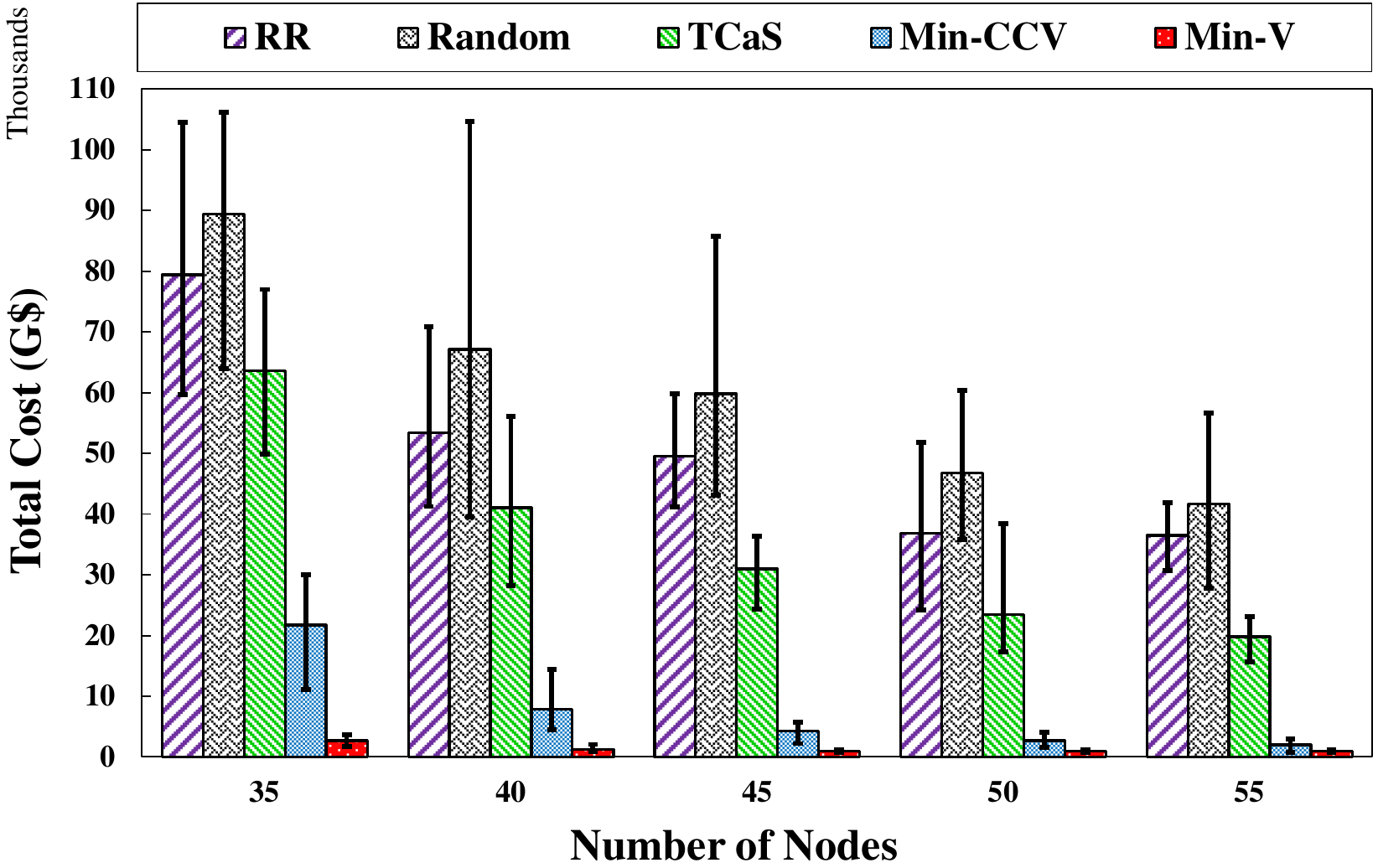}
		\caption{Total cost}
		\label{fig:exp3-Total}
	\end{subfigure}
	\caption{Simulation results for \textit{Experiment 3} (impact of varying number of cloud nodes).}
	\label{fig:fig6}
\end{figure}

\textbf{3) Impact of a varying number of cloud nodes:} As the final experiment, we focus on changing the number of cloud nodes from 5 to 25. The simulation results for this experiment reveal that Min-V and then Min-CCV perform better than the rest in terms of PDST, violation cost, computation cost, and total cost (see Figs.~\ref{fig:fig6}a to ~\ref{fig:fig6}f). Here there are some important points to mention. First, as the number of cloud nodes reaches from 5 to 15, the PDST of the both proposed algorithms substantially improves, and their violation cost significantly reduces (see Figs.~\ref{fig:fig6}a and ~\ref{fig:fig6}c). However, after that, we observe a little impact on the performance of the system. The main reason behind this is that the 15 cloud nodes are enough to process the delay-tolerant tasks, but the latency-sensitive tasks cannot meet their deadline using cloud nodes. Again, as Fig.~\ref{fig:fig6}b shows, our algorithms and TCaS give better results than RR and Random concerning makespan. For the computation cost aspect, our Min-V and then Min-CCV provide the lowest cost (Fig.~\ref{fig:fig6}b) while for the communication cost, RR and then Random provides the most economical cost (Fig.~\ref{fig:fig6}e). Fig.~\ref{fig:fig6}f depicts that Min-V, Min-CCV and TCaS give the first, second, and third best total cost, respectively.
 
\section{Discussions}\label{discuss}

Despite the effectiveness of our proposed algorithms, several aspects are remaining that we address them in the following.  First, similar to \cite{8,18}, the presented work does not consider the dependency between tasks on the application/job level. Given that in the real world, the tasks of some applications are interdependent, the proposed algorithms can be extended in a way that they work for such applications too. Second, in a Software-Defined Networking (SDN)-enabled volunteer system, computing devices are usually deployed by different operators and owners in a geographically distributed environment \cite{luan2015fog}. Hence, this introduces a challenge to the fog broker in terms of data routing among fog devices. However, in this work, we have ignored this aspect. Therefore, an effective routing protocol should be integrated into our algorithms to cope with this issue. Finally, due to the low time and space complexity and the high performance of the proposed methods, we can use them in the online task scheduling problem in the domain of fog-cloud computing.

\section{Conclusion and Future Directions}\label{conc}
In this paper, we focused on the task scheduling problem in volunteer fog-cloud environments. We formulated the problem as mixed-integer programming to minimize computation, communication and violation costs. To address the problem, two efficient heuristic algorithms are introduced. The proposed algorithms were evaluated in terms of their performance via various experiments. The results show that our algorithms significantly outperform others in terms of percentage of deadline satisfied tasks (PDST) and violation cost. They also provide low makespan and computation cost. Specifically, the proposed algorithms are capable of delivering PDST by up to 95\% and reducing the violation cost by up to 99.5\%.

 As the number and scale of volunteer computing systems are growing rapidly, the energy consumed by their computing resources is significantly increasing, which imposes a substantial impact on the system cost. Therefore, as future work, we plan to expand the scheduling problem by considering energy consumption optimization is another key metric. Moreover, to further assess the efficiency of the proposed algorithms, we would like to evaluate their performance under different real data sets and compare them with other state-of-the-art. Last but not least, this work can be extended in future Internet technologies, including edge systems. Designing an automatic resource allocation mechanism is one of the prominent goals of the network brokers, especially brokers who are dealing with the real-time and sensitive applications like video streaming used for remote surgery. Hence, we plan to add artificial intelligence methods to our heuristic approaches which shape a zero-tough model mimics the environment changes. 

	
\bibliographystyle{ACM-Reference-Format}

\bibliography{main}

\end{document}